\documentclass[aps,twocolumn,showpacs,showkeys, superscriptaddress,floatfix]{revtex4}

\usepackage{graphicx}
\usepackage[utf8]{inputenc}
\usepackage{amsmath}
\usepackage{mathtools}
\usepackage{verbatim}
\usepackage{amssymb}
\usepackage{color}
\usepackage{appendix}
\usepackage{subfigure}
\usepackage{tikz}
\usepackage{ifthen}
\usepackage{hyperref}
\usepackage{algorithm}
\usepackage{algpseudocode}
\usepackage{tabularx}
\usepackage{amsmath, amssymb}
\usepackage{makecell}
\usepackage{multirow}
\hypersetup{
    colorlinks=true,
    linkcolor=blue,
    filecolor=gray,
    urlcolor=blue,
    citecolor=blue,
    }
\usetikzlibrary{calc,fadings,decorations.pathreplacing}
\usepackage[normalem]{ulem}

\newcommand{\cone}{orange}
\newcommand{\ctwo}{blue}

\newcommand{\cfour}{cyan}

\newcommand{\csix}{olive}
\newcommand{\cseven}{purple}

\newcommand{\SVO}[1]{
\begin{tikzpicture}[scale=#1]
\clip (-0.15,-0.13) rectangle (0.16,0.18);
\draw [line width=0.5,black] (-0.07,0.17)-- (0.14,0.17);
\node at (0,0) {\ifthenelse{\equal{#1}{1}}{$\vec{S}$}{{\scriptsize $\Svec$}}}; \end{tikzpicture}
}

\newcommand{\SVT}[1]{
\begin{tikzpicture}[scale=#1]
\clip (-0.15,-0.13) rectangle (0.16,0.23);
\draw [line width=0.5,black] (-0.07,0.16)-- (0.14,0.16);
\draw [line width=0.5,black] (-0.07,0.22)-- (0.14,0.22);
\node at (0,0) {\ifthenelse{\equal{#1}{1}}{$\vec{S}$}{{\scriptsize$\Svec$}}}; \end{tikzpicture}
}

\newcommand{\KO}[1]{\begin{tikzpicture}[scale=#1]
\clip (-0.15,-0.13) rectangle (0.16,0.18);
\draw [line width=0.5,black] (-0.08,0.18)-- (0.15,0.18);
\node at (0,0)  {\ifthenelse{\equal{#1}{1}}{$K$}{{\scriptsize $K$}}}; \end{tikzpicture}}

\newcommand{\KT}[1]{\begin{tikzpicture}[scale=#1]
\clip (-0.15,-0.13) rectangle (0.16,0.22);
\draw [line width=0.5,black] (-0.10,0.16)-- (0.15,0.16);
\draw [line width=0.5,black] (-0.10,0.22)-- (0.15,0.22);
\node at (0,0)  {\ifthenelse{\equal{#1}{1}}{$K$}{{\scriptsize $K$}}};\end{tikzpicture}}

\newcommand{\GT}[1]{
\begin{tikzpicture}[scale=#1]
\clip (-0.15,-0.13) rectangle (0.15,0.25);
\draw [line width=0.5,black] (-0.14,0.16)-- (0.14,0.16);
\draw [line width=0.5,black] (-0.14,0.21)-- (0.14,0.21);
\node at (0,0) {\ifthenelse{\equal{#1}{1}}{$G$}{{\scriptsize $G$}}}; \end{tikzpicture}
}

\newcommand{\WT}[1]{
\begin{tikzpicture}[scale=#1]
\clip (-0.15,-0.08) rectangle (0.15,0.20);
\draw [line width=0.5,black] (-0.12,0.12)-- (0.11,0.12);
\draw [line width=0.5,black] (-0.12,0.17)-- (0.11,0.17);
\node at (0,0) {\ifthenelse{\equal{#1}{1}}{$\omega$}{{\scriptsize $\omega$}}}; \end{tikzpicture}
}

\newcommand{\CT}[1]{
\begin{tikzpicture}[scale=#1]
\clip (-0.13,-0.128) rectangle (0.13,0.20);
\draw [line width=0.5,black] (-0.12,0.15)-- (0.11,0.15);
\draw [line width=0.5,black] (-0.12,0.20)-- (0.11,0.20);
\node at (0,0) {\ifthenelse{\equal{#1}{1}}{$\mathcal{C}$}{{\scriptsize $\mathcal{C}$}}}; \end{tikzpicture}
}

\newcommand{\mo}[1]{
\begin{tikzpicture}[scale=#1]
\clip (-0.18,-0.075) rectangle (0.18,0.135);
\draw [line width=0.5,black] (-0.12,0.13)-- (0.12,0.13);
\node at (0,0) {\ifthenelse{\equal{#1}{1}}{$m$}{{\scriptsize $m$}}}; \end{tikzpicture}
}

\renewcommand\vec[1]{\boldsymbol{#1}}

\newcommand{\Svec}{\vec{S}}

\makeatletter
\newcommand{\multiline}[1]{%
  \begin{tabularx}{\dimexpr\linewidth-\ALG@thistlm}[t]{@{}X@{}}
    #1
  \end{tabularx}
}
\makeatother

\begin{document}
\date{\today}\title{Graphical Representations and Worm Algorithms for the O($N$) Spin Model}
\author{Longxiang Liu\footnote{The authors contribute equally to this paper.}}
%\email{lx\_liu@ustc.edu.cn}
\affiliation{Simulation of Physical Systems Division, Beijing Computational Science Research Center, Beijing 100193, China}
\affiliation{National Laboratory for Physical Sciences at Microscale, University of Science and Technology of China, Hefei, Anhui 230026, China}
%\affiliation{Hefei National Laboratory, University of Science and Technology of China, Hefei 230088, China}
%\affiliation{CAS Center for Excellence and Synergetic Innovation Center in Quantum Information and Quantum Physics, University of Science and Technology of China, Hefei, Anhui 230026, China}
\author{Lei Zhang\footnotemark[\value{footnote}]}
\affiliation{Department of Modern Physics, University of Science and Technology of China, Hefei, Anhui 230026, China}
%\affiliation{Hefei National Laboratory, University of Science and Technology of China, Hefei 230088, China}
\author{Xiaojun Tan}
\affiliation{Department of Modern Physics, University of Science and Technology of China, Hefei, Anhui 230026, China}
%\affiliation{Hefei National Laboratory, University of Science and Technology of China, Hefei 230088, China}
\author{Youjin Deng}
\email{yjdeng@ustc.edu.cn}
\affiliation{National Laboratory for Physical Sciences at Microscale, University of Science and Technology of China, Hefei, Anhui 230026, China}
\affiliation{Department of Modern Physics, University of Science and Technology of China, Hefei, Anhui 230026, China}

\begin{abstract}
We present a family of graphical representations for the O($N$) spin model,
where $N \ge 1$ represents the spin dimension, and $N=1,2,3$ corresponds to 
the Ising, XY and Heisenberg models, respectively. 
With an integer parameter $0 \le \ell \le N/2$, each configuration 
is the coupling of  $\ell$ copies of subgraphs consisting of directed flows
and $N -2\ell$ copies of subgraphs constructed by undirected loops,
which we call the XY and Ising subgraphs, respectively.
On each lattice site, the XY subgraphs satisfy the Kirchhoff flow-conservation law 
and the Ising subgraphs obey the Eulerian bond condition. 
Then, we formulate worm-type algorithms and simulate the O($N$) model on the simple-cubic lattice 
for $N$ from 2 to 6 at all possible $\ell$. 
It is observed that the worm algorithm has much higher efficiency than the Metropolis method, 
and, for a given $N$, the efficiency is an increasing function of $\ell$. 
Beside Monte Carlo simulations, we expect that these graphical representations 
would provide a convenient basis for the study of the O($N$) spin model 
by other state-of-the-art methods like the tensor network renormalization.
\end{abstract}
\pacs{02.70.Tt, 05.10.Ln, 05.10.-a, 64.60.De, 75.10.Hk, 75.10.Nr}

\keywords{Markov-chain Monte Carlo algorithms;  Continuous spin models; Graphical representations}
\maketitle

\section{Introduction}
The O($N$) spin model~\cite{stanley1968dependence}, also referred to as the O$(N)$ nonlinear $\sigma$ 
model in field-theoretic parlance, is central to the study of critical phenomena. 
It has been subjected to extensive studies over several decades via both field-theoretical approaches
and statistical mechanical methods. Within the comprehensive family of O$(N)$ spin models, 
the models for $N=1$, $2$, and $3$ are of particularly significant interest 
as they correspond directly to the Ising, XY, and Heisenberg models, respectively.

To surpass the constraints of perturbation theory, the powerful numerical tool—the Monte Carlo (MC)
method~\cite{binder1986monte}—is widely used to simulate the O($N$) spin model. 
However, the Metropolis scheme, which is frequently employed~\cite{metropolis1953equation}, 
undergoes significant `critical slowing down', consequently impairing computational efficiency 
near the critical point~\cite{sokal1997functional}. 
Addressing this issue, various non-local update algorithms have been put forward, 
encompassing multigrid techniques and cluster update algorithms~\cite{swendsen1987nonuniversal}. 
Among these, the cluster update algorithms prove to be the most effective for the O($N$) 
spin model~\cite{janke1996}. On the flip side, for local update algorithms, 
the worm technique emerges as a promising contender. 
This algorithm, extensively deployed in both classical and quantum 
systems~\cite{prokof2001worm,prokof1998worm}, 
exhibits superior efficiency at criticality, typically applied to graphical 
representations complying with Kirchhoff-like conservation laws. 
Fundamentally, a worm algorithm modifies configuration graphs by introducing two defects 
that infringe these conservation laws, moves them around according to 
the balance condition for MC simulations, and ultimately removes them once they reconvene. 
Hence, to develop worm-type algorithms for particular models, like the O($N$) spin model, 
it is often required to first formulate their graphical representations.

Graphical representations for the Ising ($N=1$) and XY ($N=2$) models are well established. 
For the Ising model, a specific graphical representation introduces a non-negative integer variable 
on each lattice bond, visually represented by the number of lines drawn on the corresponding 
bond (Fig.~\ref{fig:Ising1}). A configuration makes a non-zero contribution only 
when it satisfies the Eulerian condition, which requires each site to be connected 
by an even number of lines. 
On the other hand, a prevalent graphical representation for the XY model 
assigns two types of non-negative variables to each bond, 
symbolizing currents flowing in positive and negative directions, respectively. 
These currents, in their entirety, satisfy conservation at each site, 
reflecting the Kirchhoff flow-conservation law (Fig.~\ref{fig:XY1}). 
For the sake of convenience, configurations in these two representations 
are referred to as the Ising subgraphs and the XY subgraphs in this paper, respectively.

When considering an arbitrary value of $N >2$, 
studies on graphical representations have also made significant advances in the past decade. 
In 2010, Wolff put forward a graphical representation defined by a single set of bond variables, 
depicted by lines on the corresponding bonds~\cite{wolff2010simulating}. 
Each site incorporates a type of switch-board, creating pairwise connections among all neighboring lines. 
This approach implies that a range of configurations can be attributed to 
one specific set of bond variables, necessitating the inclusion of symmetry factors. 
A few years later, two distinct representations—each depicted by $N$ sets of bond variables-were proposed. 
These representations view configurations as the combination of multiple subgraph 
copies~\cite{bruckmann2015dual,katz2017comparison}. Specifically, a configuration in Ref.\cite{bruckmann2015dual} is the union of 
one XY graph copy and $N-2$ Ising graph copies. Conversely, a configuration in Ref.\cite{katz2017comparison} is the combination of $N$ Ising graph copies.

In this work, we devise a systematic series of graphical representations for the O$(N)$ spin model 
using a streamlined approach. This series is parameterized by an integer $\ell$ ($0 \le \ell \le N/2$). 
To establish these representations, we dissect the total spin vector into 
a weighted sum of $\ell$ copies of XY vectors and $N-2\ell$ copies of Ising vectors. 
By incorporating currents (flows) for the XY vectors and undirected bonds for the Ising vectors, 
we obtain the graphical representation after integrating out the spin degree of freedom. 
The resulting configuration consists of $\ell$ copies of XY subgraphs and $N-2\ell$ copies of 
Ising subgraphs. Notably, when $\ell=0$ and $1$, our representations reduce to those proposed 
in Ref.~\cite{bruckmann2015dual,katz2017comparison}. 
Further, by introducing two defects in these graphical representations, 
we develop corresponding worm algorithms for any arbitrary $N$ and potential $\ell$. 
Using these algorithms, we conduct large-scale simulations on a simple-cubic lattice 
with a linear size of up to $L=96$ for $N=2$, $3$, $4$, $5$, and $6$, 
across all possible $\ell$ values. 
We deduce and compare the dynamic exponents for each variant of the worm algorithms from 
the fitting of the integrated autocorrelation time for an `energy-like' observable. 
Our simulation results reveal that the algorithm efficiency increases with the number 
of XY subgraph copies in a single configuration, i.e., a larger $\ell$ promotes higher efficiency. 
In the best cases, dynamic exponents can reach $z=0.20(1)$, $0.32(2)$, $0.22(1)$, $0.26(3)$, 
and $0.16(2)$ for $N=2$, $3$, $4$, $5$, and $6$, respectively. 
Consequently, the efficiency of worm-type algorithms is comparable with 
that of cluster update methods~\cite{Deng2007SW}, 
significantly outperforming the Metropolis algorithm which has a dynamic exponent $z \approx 2$.

In addition to MC simulations, researchers are exploring other advanced numerical methods for 
applications on the O($N$) spin model, among which the tensor network 
renormalization (TNR)~\cite{evenbly2015} is one of the most promising. 
It starts by representing the partition function or the ground state wavefunction 
as a tensor network state that is formed by taking the product of local tensors defined on the lattice sites. 
However, applying this method to the O($N$) spin model is challenging 
because each spin has infinite degrees of freedom, 
and the conventional method of constructing the local tensors fails. 
For $N=1$ and $2$, researchers proposed a novel scheme to construct the tensor representation 
using character expansions~\cite{liuyuzhi2013}. 
These representations can actually be derived by slightly reformulating the conventional 
bond-based graphical representations for the Ising and XY models. 
Therefore, we expect that the development of graphical representations 
would provide a convenient foundation for investigating the O($N$) spin model 
with TNR and other numerical methods.

This paper is structured as follows: In Section~\ref{sec:Hamiltonian}, we provide a succinct introduction 
to the Hamiltonian and the partition function for the O$(N)$ spin model, 
and then revisit the graphical representations for the Ising and XY models 
before deriving the family of graphical representations for the O$(N)$ spin model. 
The worm algorithms are introduced in Section~\ref{sec:worm alg}, followed by the presentation 
of our simulation results in Section~\ref{sec:num res}. 
We conclude the paper with a discussion in Section~\ref{sec:dis}.

\section{Hamiltonian and graphical representations} \label{sec:Hamiltonian}

The O$(N)$ spin model characterizes $N$-component spin vectors with inner-product interactions on a lattice,
as represented by the Hamiltonian:
\begin{eqnarray}
	\mathcal{H}/k_B T=-\sum_{\langle i,j \rangle}K\vec{S}_i\cdot\vec{S}_j,
\end{eqnarray}
where $\vec{S}_i$ is the $N$-dimensional unit spin vector on site $i$, 
explicitly expressed as $\vec{S}=\sum_{\mu=1}^{N}S_{\mu}\vec{e}_{\mu}$ with $\vec{e}_{\mu}$ the 
unit vector along the $\mu$ direction, 
subject to the constraint $\sum_{\mu=1}^{N}S_\mu^2=1$. 
The symbol $K>0$ denotes the ferromagnetic coupling strength for each nearest neighbor pair, 
$k_B$ is the Boltzmann factor and $T$ represents the system's temperature.
The symbol $\langle i,j \rangle$ indicates that the summation covers all nearest neighboring sites. 
The corresponding partition function is given by:
\begin{equation}\label{eq:partition_function_ON spin}
\mathcal{Z}=\left(\prod_{i}\int\mathcal{D}\vec{S}_i\right)\prod_{\langle i,j \rangle}e^{K\vec{S}_i\cdot\vec{S}_j}.
\end{equation}
Here, the integral $\int\mathcal{D}\vec{S}$ signifies the uniform measure of the vector $\vec{S}$, 
defined as $\int \mathcal{D}\vec{S}=\int_{|\vec{S}|=1}d^N \vec{S}$, 
with $N$ denoting the dimension of the vector. 
Notice that, to maintain the elegance of partition function expressions, 
we always omit trivial factors, which poses no physical implications.

In the following parts, we begin by providing a brief overview of the well-established graphical 
representations for the Ising ($N=1$) and XY ($N=2$) models. 
Using these as a basis, we then construct a series of graphical representations 
adaptable to any given value of $N$. Note that, for the sake of simplicity, our illustrations 
in this section are confined to the square lattice. 
However, our derivation is applicable to any type of lattice in any spatial dimension.

\subsection{N=1: the Ising model \label{sec:Ising GR}}

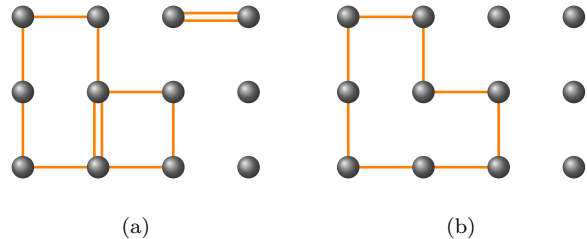
\begin{figure}[t]
	\centering
	\subfigure[]{\label{fig:Ising1}
		\begin{tikzpicture}[scale=1]
			\clip (-0.5,0.5) rectangle (3.5,3.5); % Clips the picture...			
			\draw [line width=1,\cone] (0,3)--(0,1)--(2,1)--(2,2)--(1,2)--(1,3)--(0,3);
			
			\foreach \x in {0.05} {
				\draw [line width=1,\cone] (1-\x,1)--(1-\x,2);
				\draw [line width=1,\cone] (1+\x,1)--(1+\x,2);
				%\draw [line width=1,\cone] (3-\x,2)--(3-\x,3);
				%\draw [line width=1,\cone] (3+\x,2)--(3+\x,3);
				\draw [line width=1,\cone] (2,3-\x)--(3,3-\x);
				\draw [line width=1,\cone] (2,3+\x)--(3,3+\x);
			}
			
			\foreach \x in {0,...,3} {
				\foreach \y in {0,...,3} {
					\shade[ball color=gray] (\x,\y) circle(0.15);
				}
			}
		\end{tikzpicture}
	}
	\subfigure[]{\label{fig:Ising2}
		\begin{tikzpicture}[scale=1]
			\clip (-0.5,0.5) rectangle (3.5,3.5); % Clips the picture...
			%\draw[style=help lines] (0,0) grid[step=1cm] (3,3);
			%\draw [line width=1,\cone] (1,1)--(1,0)--(0,0)--(0,1)--(1,1);
			\draw [line width=1,\cone] (0,3)--(0,1)--(2,1)--(2,2)--(1,2)--(1,3)--(0,3);
			
			\foreach \x in {0,...,3} {
				\foreach \y in {0,...,3} {
					\shade[ball color=gray] (\x,\y) circle(0.15);
				}
			}
		\end{tikzpicture}
	}
	\caption{Typical graphical configurations for the Ising model in 
	representation~(\ref{eq:Z_graphic_Ising}) (a) and 
	commonly used representation~(\ref{eq:Z_graphic_Ising2}) (b). 
	In both representations, a graph is constructed using undirected closed loops, 
	with the number of lines on each bond representing the bond variable $n_{ij}$. 
	Each of these lines contributes a weight factor $W_{\text{Is}}(n_{ij})$ or 
	$\widetilde{W}_{\text{Is}}(n_{ij})$. Notably, while representation (b) 
	allows a maximum of one line per bond, representation (a) can accommodate 
	multiple lines on a single bond.}
	\label{fig:Ising}
\end{figure}

The Ising model characterizes unit spin vectors on the lattice that can either point upwards or downwards. 
Consequently, its partition function is given by
\begin{eqnarray} \label{eq:Z_Ising}
	\mathcal{Z}_{\text{Is}}=\sum_{\{s_{i}\}}\prod_{\langle i,j \rangle}e^{Ks_is_j},
\end{eqnarray}
where $\!s_i\!=\!\pm 1\!$ signifies the spin vector at site $i$. 
The notation $\{\cdot\}$ suggests that the index inside the brackets iterates over all possible values. 
By employing a Taylor-series expansion to the Boltzmann statistical factor, 
we obtain $\exp(Ks_is_j)\!=\!\sum_{n_{ij}=0}^{+\infty}{K^{n_{ij}}\!(s_is_j)^{n_{ij}}}\!/{n_{ij}\!}$, 
where $n_{ij}=0,1,2,...$ is the variable defined on each bond. 
This expansion gives us a new form for the partition function:
\begin{eqnarray} \label{eq:Z_Ising_1}
	\!\mathcal{Z}_{\text{Is}}\!=\!\sum_{\{n_{ij}\}}\!\left(\prod_{\langle i,j \rangle} \frac{K^{n_{ij}}}{n_{ij}!}\right)\!\!\left(\prod_i\!\!\sum_{s_i=\pm1}\!\!s_i^{\sum_{j\in \mathbb{E}(i)} n_{ij}}\right)\: , 
\end{eqnarray}
where $\mathbb{E}(i)$ signifies the set of nearest neighboring sites of site $i$. 
As depicted in Fig. \ref{fig:Ising1}, if we draw a number of {\em undirected} lines on each bond 
equivalent to the corresponding bond variable $n_{ij}$, we get a graph for every summation term 
in the partition function Eq.~(\ref{eq:Z_Ising_1}), 
each with a different number of lines on each bond. 
The latter factor in each summation term implies that only 
when $n_{{ij}}$ fulfills the constraint at any site $i$:
\begin{eqnarray}
 {\cal L}_i \equiv \text{mod}\left(\sum_{j\in \mathbb{E}(i)}n_{ij},\; 2 \right)=0,
\end{eqnarray}
will the entire summation term make a non-zero contribution. This requirement implies that 
the degree of each site in the graph should be even, or alternatively, 
the graph can be seen as composed of closed loops. 
These non-zero contribution graphs are termed valid configurations, 
and if we use $G_{\text{Is}}$ to denote all of them, the partition function can be written 
in a graphical form:
\begin{equation}\label{eq:Z_graphic_Ising}
	\mathcal{Z}_{\text{Is}}=\sum_{G_{\rm Is}}\prod_{\langle i,j \rangle}W_{\text{Is}}(n_{ij}),
\end{equation}
where $W_{\text{Is}}(n)\equiv K^{n}/n!$ stands for the weight factor contributed by each bond.

Consequently, we construct a graphical representation for the Ising model, 
in which configurations are composed of closed-loop graphs founded on the bond framework. 
In this configuration graph, each bond imparts a weight factor $W_{\text{Is}}(n)$, 
dictated by the number of lines, $n$, present on it.

The above graphical representation differs from the commonly employed one for the Ising model, 
as depicted in Fig. \ref{fig:Ising2}. In the conventional representation, 
the constraint of even degree at each site still applies, but the bond variables $n_{ij}$ 
can only adopt two values: $0$ or $1$. In graphical terms, 
loops in the graph should not share any bonds or traverse a bond more than once. 
This results in the following expansion for the partition function:
\begin{equation} \label{eq:Z_graphic_Ising2}
	\mathcal{Z}_{\text{Is}}=\sum_{\widetilde{G}_{\text{Is}}}\prod_{\langle i,j \rangle}
	\widetilde{W}_{\text{Is}}(n_{ij}),
\end{equation}
Here, $\widetilde{G}_{\text{Is}}$ represents all graphs conforming to the above criteria, 
and $\widetilde{W}_{\text{Is}}(n)\equiv {\tanh^{n}K}$ is the associated weight factor for each bond. 
Indeed, we can derive this representation by summing over all graphs in 
representation~(\ref{eq:Z_graphic_Ising}) that maintain the same parity of line numbers on each bond. 
Alternatively, if we shift our focus from the bonds to the sites, we can express the partition function of the Ising model in a tensor form as follows: 
 \begin{equation} \label{eq:Z_tensor_Ising}
 	\mathcal{Z}_{\text{Is}}=\sum_{\{n_{ij}\}}\prod_{i}{\mathbb{T}_\text{Is}}^{(i)},
 \end{equation}
where ${\mathbb{T}_\text{Is}}^{(i)}$ is a tensor on site $i$, 
given by ${\mathbb{T}_\text{Is}}^{(i)}=\prod_{j\in \mathbb{E}(i)}
\sqrt{\widetilde{W}_{\text{Is}}(n_{ij})}\delta(\text{mod}(\sum_{j \in \mathbb{E}(i)}n_{ij},2))$. 
Notably, this tensor representation is consistent with that proposed in Ref.~\cite{liuyuzhi2013} 
for the Ising model.

\subsection{N=2: the XY model}\label{sec:XY GR}

\begin{figure}[t]
	\centering
	\subfigure[]{\label{fig:XY1}
		%\begin{minipage}[b]{0.3\textwidth}
		\begin{tikzpicture}[scale=1]
			\clip (-0.5,-0.5) rectangle (3.5,2.5); % Clips the picture...
			%\draw[style=help lines] (0,0) grid (3,2);
			\foreach \x in {0.62} {
				\draw [line width=1,-latex,\cfour] (0,1)-- (0,1-\x);
				\draw [line width=1,\cfour] (0,1)-- (0,0);
				\draw [line width=1,-latex,\cfour] (0,0)-- (\x,0);
				\draw [line width=1,\cfour] (0,0)-- (1,0);
				\draw [line width=1,-latex,\cfour] (1,0)-- (1,\x);
				\draw [line width=1,\cfour] (1,0)-- (1,1);
				\draw [line width=1,-latex,\cfour] (1,1)-- (1-\x,1);
				\draw [line width=1,\cfour] (1,1)-- (0,1);
				%\draw [line width=1,-latex,\cfour] (2,1)-- (2-\x,1);
				%\draw [line width=1,\cfour] (2,1)-- (1,1);
				\draw [line width=1,-latex,\cfour] (3,1)-- (3-\x,1);
				\draw [line width=1,\cfour] (3,1)-- (2,1);
				\draw [line width=1,-latex,\cfour] (1,1)-- (1,1+\x);
				\draw [line width=1,\cfour] (1,1)-- (1,2);
				\draw [line width=1,-latex,\cfour] (1,2)-- (1+\x,2);
				\draw [line width=1,\cfour] (1,2)-- (2,2);
				\draw [line width=1,-latex,\cfour] (2,2)-- (2+\x,2);
				\draw [line width=1,\cfour] (2,2)-- (3,2);
				\draw [line width=1,-latex,\cfour] (3,2)-- (3,2-\x);
				\draw [line width=1,\cfour] (3,2)-- (3,1);
				\draw [line width=1,-latex,\cfour] (2,1)-- (2-\x,1);
				\draw [line width=1,\cfour] (2,1)--(1,1);
				
				\foreach \y in {0.05} {
					\draw [line width=1,-latex,\cfour] (1,0-\y)--(1+\x,0-\y);
					\draw [line width=1,\cfour] (1,0-\y)--(2,0-\y);
					\draw [line width=1,-latex,\cfour] (2,0+\y)--(2-\x,0+\y);
					\draw [line width=1,\cfour] (1,0+\y)--(2,0+\y);
				}
			}
			
			\foreach \x in {0,...,3} {
				\foreach \y in {0,...,2} {
					
					\shade[ball color=gray] (\x,\y) circle(0.15);
				}
			}
		\end{tikzpicture}
		%\end{minipage}
	}
	\subfigure[]{
		%\begin{minipage}[b]{0.3\textwidth}
		\begin{tikzpicture}[scale=1]
			\clip (-0.5,-0.5) rectangle (3.5,2.5); % Clips the picture...
			%\draw[style=help lines] (0,0) grid (3,2);
			\foreach \x in {0.62} {
				\draw [line width=1,-latex,\ctwo] (0,1)-- (0,1-\x);
				\draw [line width=1,\ctwo] (0,1)-- (0,0);
				\draw [line width=1,-latex,\ctwo] (0,0)-- (\x,0);
				\draw [line width=1,\ctwo] (0,0)-- (1,0);
				\draw [line width=1,-latex,\ctwo] (1,0)-- (1,\x);
				\draw [line width=1,\ctwo] (1,0)-- (1,1);
				\draw [line width=1,-latex,\ctwo] (1,1)-- (1-\x,1);
				\draw [line width=1,\ctwo] (1,1)-- (0,1);
				%\draw [line width=1,-latex,\ctwo] (2,1)-- (2-\x,1);
				%\draw [line width=1,\ctwo] (2,1)-- (1,1);
				\draw [line width=1,-latex,\ctwo] (3,1)-- (3-\x,1);
				\draw [line width=1,\ctwo] (3,1)-- (2,1);
				\draw [line width=1,-latex,\ctwo] (1,1)-- (1,1+\x);
				\draw [line width=1,\ctwo] (1,1)-- (1,2);
				\draw [line width=1,-latex,\ctwo] (1,2)-- (1+\x,2);
				\draw [line width=1,\ctwo] (1,2)-- (2,2);
				\draw [line width=1,-latex,\ctwo] (2,2)-- (2+\x,2);
				\draw [line width=1,\ctwo] (2,2)-- (3,2);
				\draw [line width=1,-latex,\ctwo] (3,2)-- (3,2-\x);
				\draw [line width=1,\ctwo] (3,2)-- (3,1);
				\draw [line width=1,-latex,\ctwo] (2,1)-- (2-\x,1);
				\draw [line width=1,\ctwo] (2,1)--(1,1);
			}
			
			%\draw [line width=1,black]      (1,0)-- (2,0)-- (2,1);
			\foreach \x in {0,...,3} {
				\foreach \y in {0,...,2} {
					
					\shade[ball color=gray] (\x,\y) circle(0.15);
				}
			}
		\end{tikzpicture}
		%\end{minipage}
		\label{fig:XY2}}
	
	\caption{Typical graphical configurations for the 2D XY model in 
	representation~(\ref{eq:Z_graphic_XY}) (a) 
	and representation~(\ref{eq:Z_graphic_XY2}) (b), where the upward and rightward directions are positive. 
	In both representations, configurations comprise directed closed loops. 
	However, in representation (a), lines following both directions can be present on a single bond, 
	and $m_{ij}^{+}$ ($m_{ij}^{-}$) tallies lines along the positive (negative) direction. 
	On the other hand, in representation (b), all lines on a given bond should share the same direction, 
	and the absolute value (sign) of $m_{ij}$ is determined by the number (direction) of lines on the bond. 
	Each bond contributes a bond weight factor: $W_{xy}(m_{ij}^{\pm})$ in (a) or $\widetilde{W}_{xy}(m_{ij})$ 
	in (b). 
	}
	\label{fig:XY}
\end{figure}
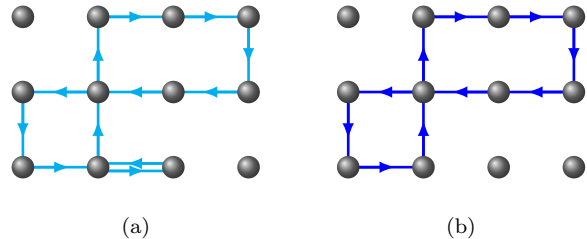

The XY model characterizes unit planar vectors at each lattice site, 
with the partition function expressed as:
\begin{eqnarray} \label{eq:Z_xy}
	\mathcal{Z}_{xy}=\left(\prod_i \int d\phi_i\right)\prod_{\langle i,j \rangle}e^{K\cos(\phi_i-\phi_j)},
\end{eqnarray}
where we have redefined the spin vector on site $i$ in terms of its angle $\phi_i\in [0,2\pi)$ 
with respect to a predetermined direction. 
By splitting the cosine function into half of the sum of $e^{i(\phi_i-\phi_j)}$ and $e^{-i(\phi_i-\phi_j)}$ 
and subsequently expanding these exponential functions into their respective 
Taylor series~\cite{prokof2001worm}, we achieve:
\begin{align}\label{eq:XY exp 1}
e^{K\!\cos(\phi_i-\phi_j)} \!=\!\!\sum_{m_{ij}^{\pm}=0}^{+\infty}\!\!\frac{(\!\frac{K}{2})^{m_{ij}^{+}\!+m_{ij}^{-}}}
{m_{ij}^{+}! \, m_{ij}^{-}!} e^{i(m_{ij}^+\!-m_{ij}^-)(\phi_i\!-\phi_j)}\!.\!
\end{align}
After integrating over all the angle variables $\phi_i$, we can represent the partition function as:
\begin{align}
	\!\!\!\!\mathcal{Z}_{xy} \!= \!\!\sum_{\{m_{\!i\!j\!}^{\!\pm}\}}\! \left(\!\prod_{\langle i,j \rangle}
	\frac{(\!\frac{K}{2}\!)^{\!m_{\!i\!j}^{\!+}\!+m_{\!i\!j}^{\!-}}}{m_{\!i\!j}^{\!+}!\, m_{\!i\!j}^{\!-}!}\!\!\right)
	%\prod_i \!\!\delta(\!\!\sum_{j\in \mathbb{E}(i)} \!\!\!\text{sgn}(i\!\!\rightarrow\!\! j)(m_{\!i\!j}^{\!+}\!\!-\!m_{\!i\!j}^{\!-})\!)\!,
 \prod_i \delta({\cal F}_i) \; ,
\end{align}
with 
\begin{align}
{\cal F}_i =\sum_{j\in \mathbb{E}(i)} \!\!\!\text{sgn}(i\!\!\rightarrow\!\! j)(m_{\!i\!j}^{\!+}\!\!-\!m_{\!i\!j}^{\!-})\; ,
\nonumber 
\end{align}
where the integer bond variables $m_{ij}^{\pm}$ can take values of $0,1,2,...$, 
and ${\cal F}_i$ represents the overall out-flow of site $i$. 
The symbol $\text{sgn}(i\!\rightarrow \!j)$ equals either $+1$ or $-1$ 
based on whether $i\!\rightarrow \!j$ is in line with or against the chosen positive direction 
for each direction (for example, we opt for the upward and rightward directions in a 2D square lattice, 
as demonstrated in Fig.~\ref{fig:XY}). 
 In contrast to the Ising model, we draw directed lines on each bond, as illustrated in Fig.~\ref{fig:XY1}. 
These lines are categorized into two types, based on whether their directions are positive or negative. 
We ensure that the numbers of both types are equal to $m_{ij}^+$ (positive) and $m_{ij}^-$ (negative), 
respectively. The Dirac $\delta$ functions in the partition function, 
which stem from the integration of $\phi_i$'s, indicate that for any site $i$, we have
\begin{equation}\label{eq:XY conser}
	{\cal F}_i \equiv \sum_{j\in \mathbb{E}(i)} \!\!\!\text{sgn}(i\!\!\rightarrow\!\! j)(m_{\!i\!j}^{\!+}\!\!-\!m_{\!i\!j}^{\!-}) = 0.
\end{equation}
This implies that the combination of positive- and negative-directed lines can be effectively treated 
as a conserved current (flow) defined on the bond structure. 
Consequently, we can represent valid configurations as graphs composed of directed closed loops. 
Furthermore, in these graphs, each bond can have multiple lines along both directions simultaneously. 
Let $G_{xy}$ denote all such graphs, the partition function $\mathcal{Z}_{xy}$ can then be expressed as:
\begin{align}\label{eq:Z_graphic_XY}
	\mathcal{Z}_{xy}=\sum_{G_{xy}}\prod_{\langle i,j \rangle}W_{xy}(m_{ij}^{\pm}),
\end{align}
where $W_{xy}(m^{\pm})\equiv (K/2)^{m^++m^-}/(m^+!m^-!)$ is the bond weight factor. This leads to a graphical representation in which configurations are graphs made up of directed closed loops. Each bond can be traversed simultaneously from both directions, contributing a weight factor $W_{xy}(m^+,m^-)$. The weight factor depends solely on the numbers of lines along the positive and negative directions, or $m^+$ and $m^-$, on the bond.

Indeed, by applying the substitution $m_{ij}=m_{ij}^+-m_{ij}^-$ and $\mo{1}_{ij}=\min\{m_{ij}^+,m_{ij}^-\}$, 
and subsequently summing over all $\mo{1}_{ij}$ variables, 
we can transform the graphical representation into a more common form:
\begin{equation}\label{eq:Z_graphic_XY2}
	\mathcal{Z}_{xy}=\sum_{\widetilde{G}_{xy}}\prod_{\langle i,j \rangle} \widetilde{W}_{xy}(m_{ij})\; ,
\end{equation}
where the bond variable $m_{ij}$ is an integer taking values $0,\pm 1,\pm 2,...$, the bond weight 
factor $\widetilde{G}_{xy}(m)$ is defined as $I_{m}(K)$, 
corresponding to the modified Bessel function of the first kind with order $m$, 
and $\widetilde{G}_{xy}$ denotes all graphs that also consist of directed closed loops. 
However, it imposes a constraint that a bond can only be traversed from a single direction at a time, 
implying that all lines on the same bond should follow the same direction: 
either positive or negative. In this representation, the quantity and direction of lines on 
a bond represent the absolute value and the sign of $m_{ij}$, respectively. 
A typical graphical configuration is illustrated in Fig.~\ref{fig:XY2}. 
Following a similar approach to the Ising model, we can apply a reformulation of 
the graphical representation in a tensor form as follows:
\begin{equation}\label{eq:Z_tensor_XY}
	\mathcal{Z}_{xy}=\sum_{\{m_{\!i\!j\!}^{\!\pm}\}}\prod_{i}{\mathbb{T}_{xy}}^{(i)}\; ,
\end{equation}
where ${\mathbb{T}_{xy}}^{(i)}\!\!=\!\!\prod_{j\in \mathbb{E}(i)}
\sqrt{\widetilde{W}_{xy}(m_{ij}^{\pm})}\delta(\text{sgn}(i\!\!\rightarrow\!\! j)
(m_{ij}^{+}-m_{ij}^{-}))$ is a tensor defined on site $i$. 
This tensor representation is identical to the one introduced in Ref.~\cite{liuyuzhi2013} for the XY model.

\subsection{Arbitrary $N$ \label{sec:O(N) spin model}}

To obtain graphical representations for any integer $N \geq 1$, we first analyze the structure of the $N$-dimensional 
spin vector $\vec{S}$ and discover that it can be decomposed into a sum of $\ell$ copies of 
two-dimensional (2D) and $N-2\ell$ copies of one-dimensional (1D) orthogonal spin vectors, 
with $0\le\ell\le N/2$. This can be expressed as:
\begin{eqnarray} \label{eq:S}
	\vec{S}&=&\sum_{\alpha}\mathcal{A}_\alpha \vec{S}^{(\alpha)}+\sum_{\beta}\mathcal{B}_\beta \vec{s}^{(\beta)}\; ,
\end{eqnarray}
where $\mathcal{A}_\alpha\vec{S}^{(\alpha)}=S_{2\alpha-1}\vec{e}_{2\alpha-1}+S_{2\alpha}\vec{e}_{2\alpha}$ 
($\alpha=1,2,...,\ell$) and $\mathcal{B}_{\beta}\vec{s}^{(\beta)}=S_{\beta+2\ell}\vec{e}_{\beta+2\ell}$ 
($\beta=1,2,...,N-2\ell$) represent the decomposed 2D or 1D spin vectors. 
The non-negative coefficients $\mathcal{A}_{\alpha}$ or $\mathcal{B}_{\beta}$ indicate the magnitude, 
while the unit vectors $\vec{S}^{(\alpha)}$ in 2D or $\vec{s}^{(\beta)}$ in 1D represent the orientations
($|\vec{S}^{(\alpha)}|=1$ and $|\vec{s}^{(\beta)}|=1$).
 Notably, the constraint for the spin vector's unitarity becomes a requirement for the coefficients:
\begin{eqnarray} \label{eq:constraintA}
	\sum_{\alpha}\mathcal{A}_{\alpha}^2+\sum_{\beta}\mathcal{B}_{\beta}^2=1,
\end{eqnarray}
or more concisely written as $\left|\vec{\mathcal{A}}\right|=1$, 
if we define an $(N-\ell)$-dimensional vector of all these magnitude 
coefficients $\vec{\mathcal{A}}\equiv(\mathcal{A}_1,...,\mathcal{A}_{\ell},\mathcal{B}_{1},...,\mathcal{B}_{N-2\ell})$.

Given that all these decomposed low-dimensional vectors $\vec{S}^{(\alpha)}$ and $\vec{s}^{(\beta)}$ are 
orthogonal to each other, the interaction between two spin vectors can be reduced to this decomposition:
\begin{eqnarray}
	K\vec{S}_i\!\cdot\!\vec{S}_j
	\!=\!\sum_{\alpha}\!K_{ij}^{(\!\alpha)}\!\cos(\phi_i^{(\alpha)}\!-\!\phi_j^{(\alpha)}\!)\!+\!\sum_{\beta}\!J_{ij}^{(\beta)}\!s_i^{(\beta)}\!s_j^{(\beta)}\!. \label{eq:S_iS_j}
\end{eqnarray}
Here, $K_{ij}^{(\alpha)}\equiv K\mathcal{A}_{i,\alpha}\mathcal{A}_{j,\alpha}$ and 
$J_{ij}^{\beta}\equiv K\mathcal{B}_{i,\beta}\mathcal{B}_{j,\beta}$, 
where $\mathcal{A}_{i,\alpha}$ or $\mathcal{B}_{i,\beta}$ are the corresponding 
coefficients $\mathcal{A}_{\alpha}$ or $\mathcal{B}_{\beta}$ for the spin at site $i$. 
Moreover, $\vec{S}_i^{(\alpha)}$ is expressed by its angle $\phi_i^{(\alpha)}\in [0,2\pi)$ 
relative to a chosen direction (for instance, the direction along $\vec{e}_{2\alpha-1}$), 
while $\vec{s}_i^{(\beta)}$ is also redefined by its single component $s_i^{(\beta)}=\pm 1$.

Therefore, we can perceive a spin vector as a weighted sum of several one-dimensional (1D) 
or two-dimensional (2D) unit spin vectors. 
Moreover, the interaction between two spin vectors can be seen as a superposition of 
interactions between the low-dimensional spin vectors that constitute each original spin vector 
and their corresponding components in the other. 
The breaking down of spin vectors and interactions into these lower-dimensional components 
reminds us of the Ising and XY models, which also utilize 1D or 2D unit spin vectors 
and comparable interactions. 
It is logical to hypothesize that we can derive a representation in which each configuration 
is built from several instances of configurations specific to the Ising or XY models.

As demonstrated in the decomposition equations~(\ref{eq:S}) and (\ref{eq:S_iS_j}), 
we can fully describe the spin vectors and their interactions 
using $\mathcal{A}_{\alpha},\mathcal{B}_{\beta}\in [0,1]$ subject to the constraint~(\ref{eq:constraintA}), 
$\phi^{(\alpha)}\in [0,2\pi)$, and $s^{(\beta)}=\pm 1$.
In this depiction, the uniform measure for the spin vector $\vec{S}$ transforms to
\begin{eqnarray}
\int\!\!\mathcal{D}\vec{S}\!=\!\int\!\!\mathcal{D}\vec{\mathcal{A}}\left|\det \!J(\vec{\mathcal{A}})\right| 
\prod_{\alpha}\int \!\!d\phi^{(\!\alpha\!)}\; \prod_{\beta}\!\!\sum_{s^{(\!\beta\!)}=\pm 1},
\end{eqnarray}
where the Jacobian determinant $\left|\det \!J(\vec{\mathcal{A}})\right|=\prod_{\alpha}\mathcal{A}_\alpha$. Note that while the Jacobian determinant is generally associated with all variables including $\{\mathcal{A}_{\alpha}\}$,  $\{\phi^{(\alpha)}\}$ and $\{s^{(\beta)}\}$, in this specific case, its absolute value solely depends on $\{\mathcal{A}_{\alpha}\}$.

Based on the decomposition of the spin vector and the interaction between two spin vectors in 
Eqs.~(\ref{eq:S}) and (\ref{eq:S_iS_j}), we can reformulate the partition function as
\begin{eqnarray}
	\mathcal{Z}\!=\!\left(\prod_{i}\int\mathcal{D}\vec{\mathcal{A}}_i \left|\det J(\vec{\mathcal{A}}_i)\right|\right)
	\prod_{\alpha}\mathcal{Z}_{xy}^{(\alpha)}\prod_{\beta}\mathcal{Z}_{\text{Is}}^{(\beta)}, \label{eq:ON exp}
\end{eqnarray}
where $\mathcal{Z}_{xy}^{(\alpha)}$ and $\mathcal{Z}_{\text{Is}}^{(\beta)}$ represent 
the partition functions for the XY (Eq.~\ref{eq:Z_xy}) and Ising (Eq.~\ref{eq:Z_Ising}) models, 
respectively. In these expressions, $\phi_i$ or $s_i$ are replaced correspondingly by $\phi_i^{(\alpha)}$ 
or $s_i^{(\beta)}$, and the interaction strength $K$ is substituted 
with $K_{ij}^{(\alpha)}$ or $J_{ij}^{(\beta)}$, 
which vary across different bonds depending on the magnitude coefficients at sites $i$ and $j$.

Employing the same procedures detailed in subsections \ref{sec:Ising GR} and \ref{sec:XY GR}, 
we can expand $Z_{xy}^{(\alpha)}$ and $Z_{\text{Is}}^{(\beta)}$ to match the forms found 
in Eqs.~(\ref{eq:Z_graphic_XY}) and (\ref{eq:Z_graphic_Ising}), respectively. 
If we denote the corresponding graphical configuration sets as $G_{xy}^{(\alpha)}$ 
and $G_{\text{Is}}^{(\beta)}$, and represent the bond variables by $m_{ij}^{\pm(\alpha)}$ 
and $n_{ij}^{(\beta)}$, respectively, 
we can then illustrate the partition function in the following graphical form:
\begin{eqnarray}\label{eq:Z_O(N)}
	\mathcal{Z}\!=\!\!\sum_{G}\prod_{\langle i,j \rangle}\!\!W(\{m_{\!i\!j}^{\!\pm(\!\alpha\!)}\}\!,\{n_{\!i\!j}^{(\!\beta\!)}\})
	\prod_i \!\mathcal{I}(\{k_i^{\!(\!\alpha\!)}\},\{l_i^{(\!\beta\!)}\}).
\end{eqnarray}
Here, $G$ is the direct sum of all the graph sets $G_{xy}^{(\alpha)}$ and $G_{\text{Is}}^{(\beta)}$. 
This implies that a graph in $G$ can be perceived as a superposition of $\ell$ copies of graphs for the XY model 
and $N-2\ell$ ones for the Ising model. 
Mathematically, for any $\alpha$ or $\beta$, the bond variables must satisfy the following constraints for any site $i$:
\begin{eqnarray}\label{eq:constraint_for_total}
	\left\{
	\begin{split}
		&{\cal F}_i^{(\alpha)} \!=\!\sum_{j\in\mathbb{E}(i)}\text{sgn}(i\rightarrow j)(m_{ij}^{+(\alpha)}-m_{ij}^{-(\alpha)})=0,\\
		&{\cal L}_i^{(\beta)} \!=\!\!\! \quad\text{mod} (\!\sum_{j\in\mathbb{E}(i)} \!\!\! n_{ij}^{(\beta)},\,2)=0.
	\end{split}\right.
\end{eqnarray}
The weight factor for each bond, denoted as $W(\!\{\!m^{\!\pm(\!\alpha\!)}\!\}\!,\!\{\!n^{(\!\beta\!)}\!\}\!)$, 
is the product of all weight factors from each decomposed graph. This can be expressed explicitly as:
\begin{eqnarray} \label{eq:W}
	W(\{m^{\!\pm(\!\alpha\!)}\}\!,\{n^{(\!\beta\!)}\})\!=\!\prod_{\alpha}\!W_{xy}(m^{\!\pm(\!\alpha\!)})\!\prod_{\beta}\!W_{\text{Is}}(n^{(\!\beta\!)}),
\end{eqnarray}
where $W_{xy}$ and $W_{\text{Is}}$ are precisely as defined in Eqs.~(\ref{eq:Z_graphic_XY}) 
and (\ref{eq:Z_graphic_Ising}).
The weight factor, $\mathcal{I}(\{k_{i}^{\!(\!\alpha\!)}\},\{l_{i}^{(\!\{\!\beta\!\}\!)}\})$ 
corresponds to site $i$, and couples all copies of the decomposed graphs in a single configuration, 
irrespective of whether they pertain to the XY model or the Ising model. This factor can be expressed as:
\begin{eqnarray} \label{eq:I}
	\!\!\!\!\!\!\!\!\!\mathcal{I}(\{k_{i}^{\!(\!\alpha\!)}\},\{l_{i}^{(\!\beta\!)}\})&\!=&\!\!\int \mathcal{D}\vec{\mathcal{A}}_i\prod_{\alpha}\mathcal{A}_{i,\alpha}^{k_i^{(\alpha)}+1}\prod_{\beta}\mathcal{B}_{i,\beta}^{l_i^{(\beta)}} \nonumber\\
	&\!=&\!\!\frac{\prod_{\alpha}\Gamma(\frac{k_{i}^{(\alpha)}+2}{2})\prod_{\beta} \Gamma(\frac{l_i^{(\beta)}+1}{2})}{2\Gamma(\frac{\sum_{\alpha} k_{i}^{(\alpha)}+\sum_{\beta}l_{i}^{(\beta)}+N}{2})}\; ,
\end{eqnarray}
where 
\begin{eqnarray} 
	k_i^{(\alpha)}\!&=&\!\sum_{j\in\mathbb{E}(i)}(m_{\!i\!j}^{\!+\!(\!\alpha\!)}\!
	                       +m_{\!i\!j}^{\!-\!(\!\alpha\!)}\!) \nonumber \\	
	l_i^{(\beta)}\!&=&\!\sum_{j\in\mathbb{E}(i)}n_{\!i\!j}^{\!(\!\beta\!)}
	\label{eq:kl}
\end{eqnarray}
represent the count of all lines connecting to site $i$ in the $\alpha$th copy 
of XY graph and the $\beta$th copy of Ising graph, respectively. 
The exponents $k_i^{\!(\!\alpha\!)}$ and $l_i^{(\!\beta\!)}$ of $\mathcal{A}_{i,\alpha}$ 
and $\mathcal{B}_{i,\beta}$ in the integral arise from the substitution of $K_{ij}^{(\alpha)}$ and $J_{ij}^{(\beta)}$ 
by their respective definition equations. 
Furthermore, an additional $+1$ is present in the exponent of $\mathcal{A}_{i,\alpha}$ 
as a result of the Jacobi determinant. 
For more details about the derivation of Eq.~(\ref{eq:I}), please see Appendix A.

It is important to emphasize that if we expand $\mathcal{Z}_{xy}^{(\alpha)}$ 
and $\mathcal{Z}_{\text{Is}}^{(\beta)}$ into their more conventional forms, 
as seen in Eqs.~(\ref{eq:Z_graphic_XY2}) and (\ref{eq:Z_graphic_Ising2}), 
we won't be able to independently separate the magnitude coefficients from the bond weight factors. 
Consequently, it would not be possible to analytically integrate them out beforehand to 
calculate the site weight factors.

\begin{figure}[t]
  \centering
  \subfigure[]{
  %\begin{minipage}[b]{0.3\textwidth}
  \begin{tikzpicture}[scale=1]
    \clip (-0.5,-0.5) rectangle (3.5,3.5); % Clips the picture...
    %\draw[style=help lines] (0,0) grid (3,2);
   \draw [line width=1,\cone] (1,2)--(1,1)--(0,1)--(0,0)--(2,0)--(2,1);
   \draw [line width=1,\csix] (2,1)--(3,1)--(3,2);
   \draw [line width=1,\cseven] (3,2)--(3,3)--(1,3)--(1,2);

    \foreach \x in {0.05} {
    \draw [line width=1,\cone] (1-\x,0)--(1-\x,1);
    \draw [line width=1,\cone] (1+\x,0)--(1+\x,1);
    \draw [line width=1,\cone] (2-\x,1)--(2-\x,2);
    \draw [line width=1,\csix] (2+\x,1)--(2+\x,2);
    \draw [line width=1,\cone]   (1,2-\x)--(2,2-\x);
    \draw [line width=1,\cseven] (1,2+\x)--(2,2+\x);
    \draw [line width=1,\csix]   (2,2-\x)--(3,2-\x);
    \draw [line width=1,\cseven] (2,2+\x)--(3,2+\x);
    }

   %\draw [line width=1,black]      (1,0)-- (2,0)-- (2,1);
     \foreach \x in {0,...,3} {
     \foreach \y in {0,...,3} {

     \shade[ball color=gray] (\x,\y) circle(0.15);
     }
     }

  \end{tikzpicture}
  %\end{minipage}
  }\label{fig:O3 1}
  \subfigure[]{
  %\begin{minipage}[b]{0.3\textwidth}
  \begin{tikzpicture}[scale=1]
    \clip (-0.5,-0.5) rectangle (3.5,3.5); % Clips the picture...
    %\draw[style=help lines] (0,0) grid (3,2);
    \foreach \x in {0.62} {
    \draw [line width=1,-latex,\cfour] (1,1)--(1-\x,1);
    \draw [line width=1,\cfour] (1,1)--(0,1);
    \draw [line width=1,-latex,\cfour] (0,1)--(0,1-\x);
    \draw [line width=1,\cfour] (0,1)--(0,0);
    \draw [line width=1,-latex,\cfour] (0,0)--(\x,0);
    \draw [line width=1,\cfour] (0,0)--(1,0);
    \draw [line width=1,-latex,\cfour] (1,0)--(1+\x,0);
    \draw [line width=1,\cfour] (1,0)--(2,0);
    \draw [line width=1,-latex,\cfour] (2,0)--(2,0+\x);
    \draw [line width=1,\cfour] (2,0)--(2,1);
    \draw [line width=1,-latex,\cfour] (1,2)--(1,2-\x);
    \draw [line width=1,\cfour] (1,2)--(1,1);
    \draw [line width=1,\cone] (2,1)--(3,1)--(3,3)--(1,3)--(1,2);

    \foreach \y in {0.05}{
    \draw [line width=1,\cone]  (2,2-\y)--(3,2-\y);
    \draw [line width=1,\cone]  (2,2+\y)--(3,2+\y);
    \draw [line width=1,-latex,\cfour] (2-\y,1)-- (2-\y,1+\x);
    \draw [line width=1,\cfour] (2-\y,1)-- (2-\y,2);
    \draw [line width=1,-latex,\cfour] (2,2-\y)-- (2-\x,2-\y);
    \draw [line width=1,\cfour] (2,2-\y)-- (1,2-\y);
    \draw [line width=1,\cone] (2+\y,1)-- (2+\y,2);
    \draw [line width=1,\cone] (2,2+\y)-- (1,2+\y);
    \draw [line width=1,-latex,\cfour] (1+\y,0)-- (1+\y,0+\x);
    \draw [line width=1,\cfour] (1+\y,0)-- (1+\y,1);
    \draw [line width=1,-latex,\cfour] (1-\y,1)-- (1-\y,1-\x);
    \draw [line width=1,\cfour] (1-\y,1)-- (1-\y,0);
    }
    }

    \foreach \x in {0,...,3} {
     \foreach \y in {0,...,3} {

     \shade[ball color=gray] (\x,\y) circle(0.15);
     }
     }
  \end{tikzpicture}
  %\end{minipage}
  }\label{fig:O3 2}
\caption{Typical graphical configurations for the O(3) spin model at different $\ell$ values:
	(a) For $\ell=0$, the configuration is composed of three copies of the Ising model graphs, 
	as shown by {\cone}, {\csix}, and {\cseven} undirected lines, respectively;
	(b) For $\ell=1$, the configuration includes one copy of the XY model graph, 
	denoted by the {\cfour} directed lines, and an additional copy of the Ising model graph, 
	represented by {\cone} undirected lines.
	In both instances, each bond or site contributes a weight factor as specified in 
	Eqs.~(\ref{eq:W}) or (\ref{eq:I}), respectively.
}
\label{fig:O3}
\end{figure}
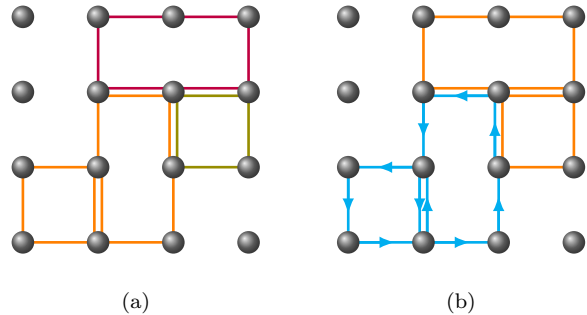

Thus, for each potential value of $\ell$, we obtain a graphical representation. 
In this representation, every graph configuration comprises $\ell$ instances of XY graphs and $N-2\ell$ 
instances of Ising graphs. Each bond and site within these configurations contribute a weight factor, 
as indicated in Eqs.~(\ref{eq:W}) and (\ref{eq:I}) respectively. 
For instance, when $N=3$, there are two viable values for $\ell$: $0$ and $1$. 
As depicted in Fig.~\ref{fig:O3}, when $\ell=0$, the graphical configuration 
can be interpreted as being comprised of three Ising graphs. 
Conversely, when $\ell=1$, the configuration can be viewed as consisting of 
a single XY graph and one Ising graph.

Furthermore, the partition function in the graphical description can be represented in tensor form 
as follows:
\begin{eqnarray}\label{eq:Z_tensor_O(N)}
	\mathcal{Z}\!=\!\!\sum_{\{m_{ij}^{\!\pm(\!\alpha\!)}\}\!,\{n_{ij}^{(\!\beta\!)}\}}\prod_{i}\mathbb{T}^{(i)},
\end{eqnarray}
where the tensor $\mathbb{T}^{(i)}$ is defined as
\begin{eqnarray}
    \mathbb{T}^{(i)}&=&\mathcal{I}(\{k_i^{\!(\!\alpha\!)}\},\{l_i^{(\!\beta\!)}\})\prod_{j\in\mathbb{E}(i)}\sqrt{W(\{m_{\!i\!j}^{\!\pm(\!\alpha\!)}\}\!,\{n_{\!i\!j}^{(\!\beta\!)}\})}\nonumber\\
    &&\cdot\prod_{\alpha} \delta({\cal F}^{(\alpha)}) \; \prod_{\beta} \delta({\cal L}^{(\beta)}) \; .
	%&&\cdot\prod_{\alpha}\delta(\sum_{j\in\mathbb{E}(i)}\text{sgn}(i\!\!\rightarrow\!\! j)(m_{ij}^{+(\alpha)}-m_{ij}^{-(\alpha)}))\nonumber\\
    %&&\cdot\prod_{\beta}\delta(\text{mod}(\sum_{j\in\mathbb{E}(i)}n_{ij}^{(\beta)},2)).
\end{eqnarray}
Consequently, each graphical representation yields a corresponding tensor representation, 
marking the initial stride towards the application of the TNR method to the O($N$) spin model.

\section{Worm Algorithm}\label{sec:worm alg}

Based on the series of graphical representations derived in the previous section, 
we formulate worm algorithms to simulate the O$(N)$ spin model for arbitrary $N$ and $\ell$. 
In these algorithms, starting from an arbitrary valid configuration, we introduce two defects 
named $Ira$ and $Masha$ into a randomly selected copy of an XY or Ising subgraph with a certain 
acceptance probability, causing the constraint in Eq.~(\ref{eq:constraint_for_total}) to fail for 
the corresponding $\alpha$ or $\beta$ of the selected subgraph at these two sites. 
In the defected subgraph, both $Ira$ and $Masha$ are connected by odd lines if it is an Ising subgraph, 
and one extra current flows in (out) for $Ira$ ($Masha$) if it is an XY subgraph. 
To update the selected subgraph, we then move the defect $Ira$ randomly by 
deleting (adding) a line or deleting (adding) a positive (negative) current, 
depending on whether it is an XY or Ising subgraph, until $Ira$ meets $Masha$ again.

We designate the space built by configurations with two defects as the `worm space'. 
Configurations within this space are weighted according to the two-point correlation function:
\begin{eqnarray}
	\mathcal{Z}(I,M)\!\!&\equiv&\!\!\langle \vec{S}_I\cdot\vec{S}_M \rangle\mathcal{Z} \nonumber\\
	&=&\sum_{\alpha}\mathcal{Z}_{xy}^{(\alpha)}(I,M)+\sum_{\beta}\mathcal{Z}_{\text{Is}}^{(\beta)}(I,M),
\end{eqnarray}
where $\vec{S}_I$ and $\vec{S}_M$ represent the spin vectors at the sites $Ira$ and $Masha$, 
respectively, and 
$\mathcal{Z}_{xy}^{(\alpha)}(I,M)\equiv\langle\vec{S}_{I}^{(\alpha)}\cdot\vec{S}_{M}^{(\alpha)}\rangle\cdot\mathcal{Z}$ 
and $\mathcal{Z}_{\text{Is}}^{(\beta)}(I,M)\equiv\langle\vec{s}_I^{(\beta)}\cdot\vec{s}_M^{(\beta)}\rangle\cdot\mathcal{Z}$ 
denote the correlation functions for the respective decomposed spin vectors. 
They display the weights of the configurations when the defects are introduced into 
the corresponding subgraphs.

Following the procedure outlined in Sec.~\ref{sec:Hamiltonian}, we can formulate the expansions 
of $\mathcal{Z}_{xy}^{(\alpha)}(I,M)$ and $\mathcal{Z}_{\text{Is}}^{(\beta)}(I,M)$ in a manner 
similar to Eq.~(\ref{eq:Z_O(N)}). These expansions follow the same bond weight factor formula 
as presented in Eq.~(\ref{eq:W}), and adopt a similar formula for the site weight factor 
as seen in Eq.~(\ref{eq:I}). In this case, $\vec{S}_{I/M}^{(\alpha)}$ or $\vec{s}_{I/M}^{(\beta)}$ 
within the angular brackets each contribute an additional $+1$ to the exponents 
of $\mathcal{A}_{I/M,\alpha}$ or $\mathcal{B}_{I/M,\beta}$ in the integration.

Due to the unitary modulus of the vector $\vec{S}$, the correlation function $\mathcal{Z}(I,M)$ 
simplifies to the partition function $\mathcal{Z}$ when $I=M$. 
Thus, we can equivalently view configurations with both $Ira$ and $Masha$ on the same site as physical 
configurations. In this context, we can measure the observables on these configurations, 
disregarding the presence of both defects.

The procedures we use to update the Ising or XY subgraphs are identical to those employed 
for the Ising or XY model. A comprehensive description of the entire algorithm for a specific $\ell$ 
graphical representation can be found in Algorithm~\ref{alg:worm}.

\begin{algorithm}[ht]
\caption{Worm Algorithm for the O($N$) spin model \label{alg:worm}}
\begin{algorithmic}[l]
	\If {$I=M$ }
        \State
        Randomly choose a site $I'$ from all sites in the lattice, and a copy of the subgraphs 
		denoted as $\gamma'$ from $\ell$ copies of XY graphs and $N-2\ell$ copies of Ising graphs.
        \State
        With the acceptance probability $P_{\gamma\rightarrow\gamma'}^{acc}(I\rightarrow I')$, 
		set $I'$ as the new location of both $Ira$ and $Masha$, and change current modified subgraph 
		to $\gamma'$, namely set $I=M=I'$, $\gamma=\gamma'$.
	\EndIf
	\If {$\gamma$ is an XY subgraph}
    	\State
       	Randomly select a site $I'$ from all the nearest neighbor sites of site $I$.
       	\State
       	If $I\rightarrow I'$ is along the positive direction we selected in advance, set $\lambda=+1$, 
		otherwise, set $\lambda=-1$.
       	\State
       	Choose $sgn$ to be $+$ or $-$ with equal probability.
       	\State
       	Set $m_{II'}^{sgn(\gamma)}=m_{II'}^{sgn(\gamma)}+\lambda\cdot sgn$ and move $Ira$ to $I'$, 
		namely set $I=I'$, with the acceptance probability $P_{\gamma,{xy}}^{acc}(I\rightarrow I',sgn)$, 
		unless this operation makes $m_{II'}^{sgn(\gamma)}$ negative and is rejected directly. 
		Here, $sgn$ stands for $+1$ or $-1$ in the calculation correspondingly.
	\Else
		\State
		Randomly select a site $I'$ from all the nearest neighbor sites of site $I$.
		\State
		Select $\lambda$ from $+1$ and $-1$ with equal probability.
		\State
		Set $n_{II'}^{(\gamma)}=n_{II'}^{(\gamma)}+\lambda$ and move $Ira$ to $I'$, namely set $I=I'$, 
		with the acceptance probability $P_{\gamma,{\rm Is}}^{acc}(I\rightarrow I')$, 
		unless this operation makes $n_{II'}^{(\gamma)}$ negative and is rejected directly.
    \EndIf
\end{algorithmic}
\end{algorithm}
\noindent The specific expressions for the acceptance probabilities, 
$P_{\gamma\rightarrow\gamma'}^{acc}(I\rightarrow I')$, $P_{\gamma,{xy}}^{acc}(I\rightarrow I',sgn)$ 
and $P_{\gamma,{\rm Is}}^{acc}(I\rightarrow I')$,
are provided in Appendix \ref{ap:acceptance_probabilities}.

\section{Numerical Simulations}\label{sec:num res}

\subsection{Setup}

To probe the dynamical critical behavior of our worm algorithms, we conduct extensive simulations 
at the critical points for each $N$ at various $\ell$ values. 
First, we introduce the observable we measure:
\begin{eqnarray}
	\mathcal{N}\!=\!\sum_{\langle i,j \rangle}\left(\sum_{\alpha} \!\left(m_{\!i\!j}^{\!+\!(\!\alpha\!)}\!+\!m_{\!i\!j}^{\!-\!(\!\alpha\!)}\!\right)\!
	+\!\sum_{\beta}\!n_{\!i\!j}^{\!(\!\beta\!)}\!\right).
\end{eqnarray}
This observable counts all the lines in the graph, irrespective of their types. 
Intriguingly, the thermodynamic average of this quantity is effectively the negative of the system's energy, 
scaled by a temperature factor $T$, that is, $E=-T\langle\mathcal{N}\rangle$. 
This relationship can be readily confirmed by evaluating the formula $E=-\partial\ln Z/\partial K$.

Based on the time series of the quantity $\mathcal{N}$ in our Monte Carlo simulations, 
we compute the integrated autocorrelation time, $\tau_{\text{int},\mathcal{N}}$, 
which is defined as:
\begin{eqnarray} \label{eq:tau}
	\tau_{{\rm int},\mathcal{N}}=\frac{1}{2}+\sum_{t=1}^{+\infty}\rho_{\mathcal{N}}(t),
\end{eqnarray}
where $t$ represents time, and $\rho_{\mathcal{N}}(t)=(\langle\mathcal{N}_0\mathcal{N}_t\rangle-\langle
\mathcal{N}\rangle^2)/\text{Var}(\mathcal{N})$ is the normalized autocorrelation function 
with variance given by ${\rm Var}(\mathcal{N})\equiv\langle\mathcal{N}^2\rangle-\langle \mathcal{N} \rangle^2$. Given that an infinite time series is unattainable, we employ an upper truncation $M$ for the summation in the calculation of $\tau_{\text{int},\mathcal{N}}$. Here, $M$ is determined through a windowing method as described in Appendix C of Ref.~\cite{madras1988pivot}:
\begin{equation} \label{eq:M}
	M=\text{min}\{m\in\mathbb{N}:m \geq c {\tau}_{\text{int},\mathcal{N}}(m)\},
\end{equation}
where $\tau_{\text{int},\mathcal{N}}(m)\equiv 1/2+\sum_{t=1}^m\rho_{\mathcal{N}}(t)$ and $c$ represents the windowing parameter.

In this study, we conduct simulations of systems at criticality for all possible values of $\ell$ 
across different $N$ ranging from $2$ to $6$ on 3D cubic lattices with various system sizes $L$, 
extending up to $96$. Periodic boundary conditions are implemented in all simulations. 
For every pair of $N$ and $L$, we perform a total of $2 \times 10^8$ measurements, 
with $L/2$ worm steps between two consecutive measurements. 
The critical couplings we use for each $N$ are detailed in Table~\ref{tab:z}.

\begin{table}
\centering
\begin{tabular}{l|l|llll}
\hline\hline
\multirow{2}*{$N$}
& \multicolumn{1}{c|}{\multirow{2}*{ $K_c$}} & \multicolumn{4}{c}{$z_{\text{int},\mathcal{N}}$} \\
	&  & $\ell=0$  & $\ell=1$ & $\ell=2$ & $\ell=3$  \\
\hline
\rule{0pt}{12pt}$2$ & $0.45416476(11)$ \!\!\cite{Xu2019} & $0.50(7)$ & $0.20(1)$ & -- & -- \\
\rule{0pt}{12pt}$3$ & $0.693003(2)$ \!\!\cite{deng2005surface} & $0.49(4) $ & $0.32(2) $& -- & --  \\
\rule{0pt}{12pt}$4$ & $0.935856(2)$\!\! \cite{deng2006bulk} & $0.43(5)$ & $0.33(2)$ & $0.22(1)$ & --   \\
\rule{0pt}{12pt}$5$ & $1.1813654(19)$\!\! \cite{fernandez2005O5} & $0.38(4)$ & $0.32(5)$ & $0.26(3)$ & -- \\
\rule{0pt}{12pt}$6$ & $1.428653(12)$\cite{jianpinglv_private_communication} & $0.35(6)$ & $0.27(2)$ & $0.22(1)$ & $0.16(2)$ \\
\hline
\end{tabular}
\caption{ Critical couplings $K_c$ we applied to simulate at for each $N$, 
and corresponding dynamical critical exponents $z_{\text{int},\mathcal{N}}$ at different $\ell$ 
which are roughly estimated via fitting the data by 
Eq.~(\ref{eq:tau_int}) at large system sizes ($L>20$).
}\label{tab:z}
\end{table}

\subsection{Results}

\begin{figure}[t]
	\centering
	\includegraphics[width=0.47\textwidth]{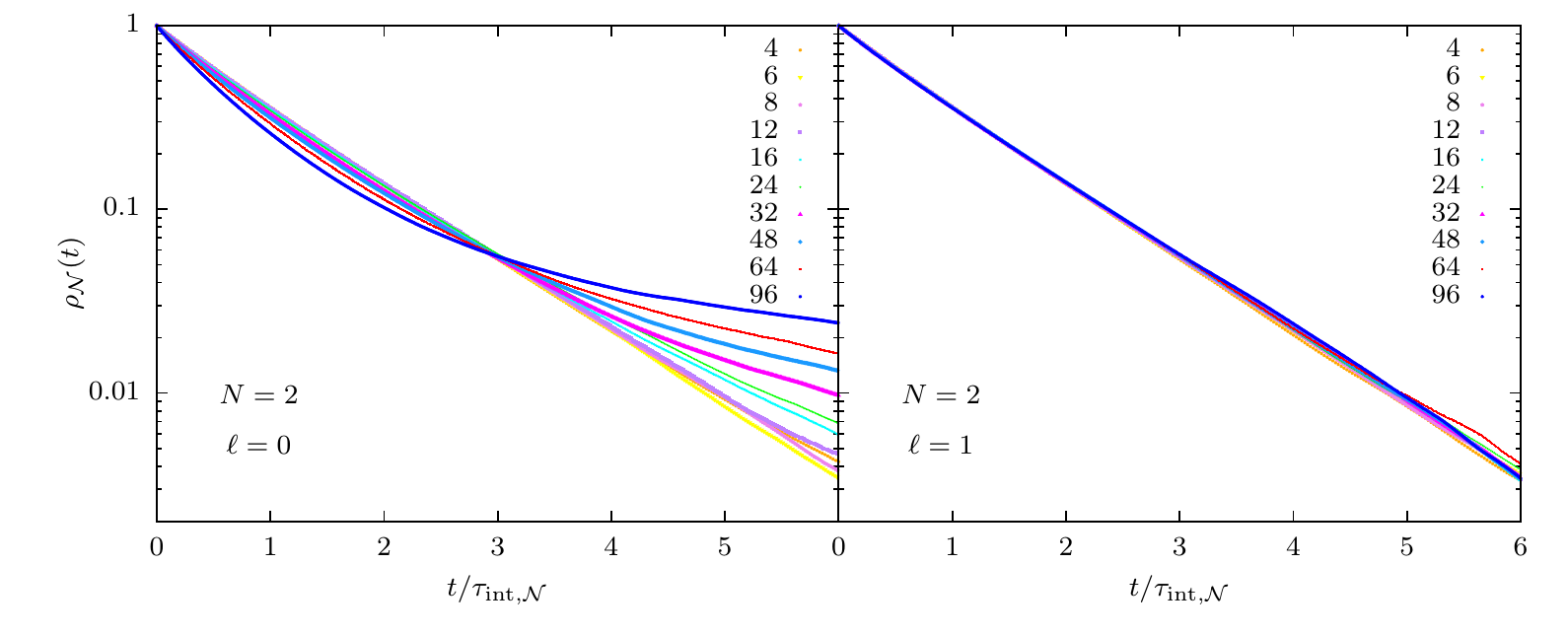}
	\caption{Autocorrelation function $\rho_{\mathcal{N}}(t)$ at criticality versus 
	$t/\tau_{\text{int},\mathcal{N}}$ for $N=2$. At $\ell=1$, $\rho_{\mathcal{N}}(t)$ 
	is very close to a pure exponential decay, while at $\ell=0$, it  exhibits a crossover 
	from a pure exponential to a complicated behavior as $L$ increases.}
	\label{fig:rhoO2}
\end{figure}

We first examine the autocorrelation function $\rho_{\mathcal{N}}(t)$ at the critical points $K_c$ 
across various system sizes $L$ for different $N$ and $\ell$ values. 
As an illustrative example, Fig.~\ref{fig:rhoO2} shows the variation of $\rho_{\mathcal{N}}(t)$ 
as a function of time $t$ for various system sizes $L$ when $N=2$. 
The figure reveals that for $\ell=1$, the decay of $\rho_{\mathcal{N}}(t)$ nearly follows 
a pure exponential across all system sizes. However, for $\ell=0$, 
while the decay is approximately exponential for smaller $L$, the trend deviates from this behavior 
as $L$ increases, leading to an intersection of lines representing $\rho_{\mathcal{N}}(t)$ 
at different $L$ around $t/\tau_{\text{int},\mathcal{N}}=3$. 
For $N=3,4,5$, and $6$, at all possible $\ell$ values, the behavior of $\rho_{\mathcal{N}}(t)$ 
is akin to that observed for $N=2$ and $\ell=0$. 
The underlying mechanism driving this phenomenon, however, remains unclear at present.
A plausible scenario is that the decay of $\rho_{\mathcal{N}}(t)$ is effectively 
the mixing of two exponential functions as $\rho_{\mathcal{N}}(t) \approx a_1 e^{-t/\tau_1}
+a_2 e^{-t/\tau_2}$. When exponents $\tau_1 > \tau_2$ and amplitudes $a_1 <a_2$, 
the decaying behavior of $\rho_{\mathcal{N}}(t)$ is dominated by $\tau_2$ for small $t$
but is eventually governed by $\tau_1$ for sufficiently large $t$.

\begin{figure}[t]
	\centering
	\includegraphics[width=0.48 \textwidth]{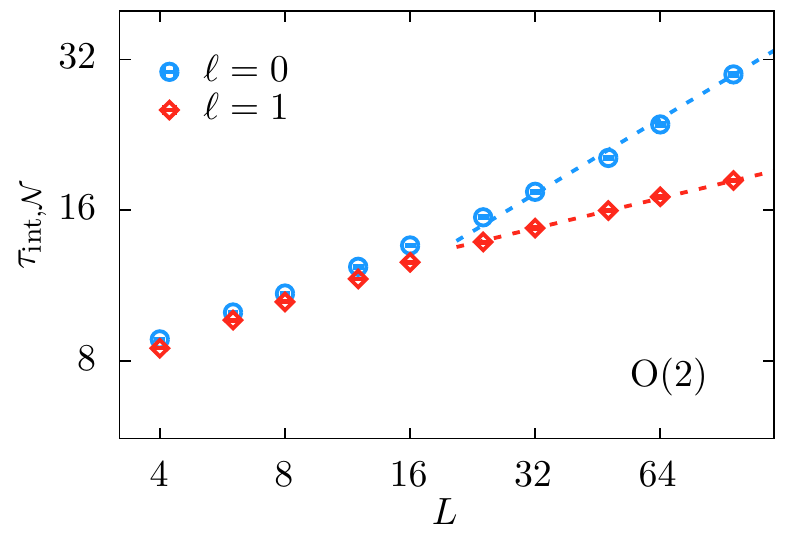}
	\caption
	{Integrated autocorrelation time $\tau_{\rm int,\mathcal{N}}$ at criticality versus system 
	size $L$ for $N=2$. Here, $\tau_{\text{int},\mathcal{N}}$ is in  the unit of one system sweep 
	which equals to the number of bonds in the system. Blue and red points are for $\ell=0$ and $1$, 
	respectively. Dashed lines show the corresponding fitting results at large $L$ with the fitting 
	formula Eq.~(\ref{eq:tau_int}). The ranges of them also mark the points we applied in the fittings.}
	\label{fig:tauO2}
\end{figure}

\begin{figure}[htb]
	\centering
	\includegraphics[width=0.48\textwidth]{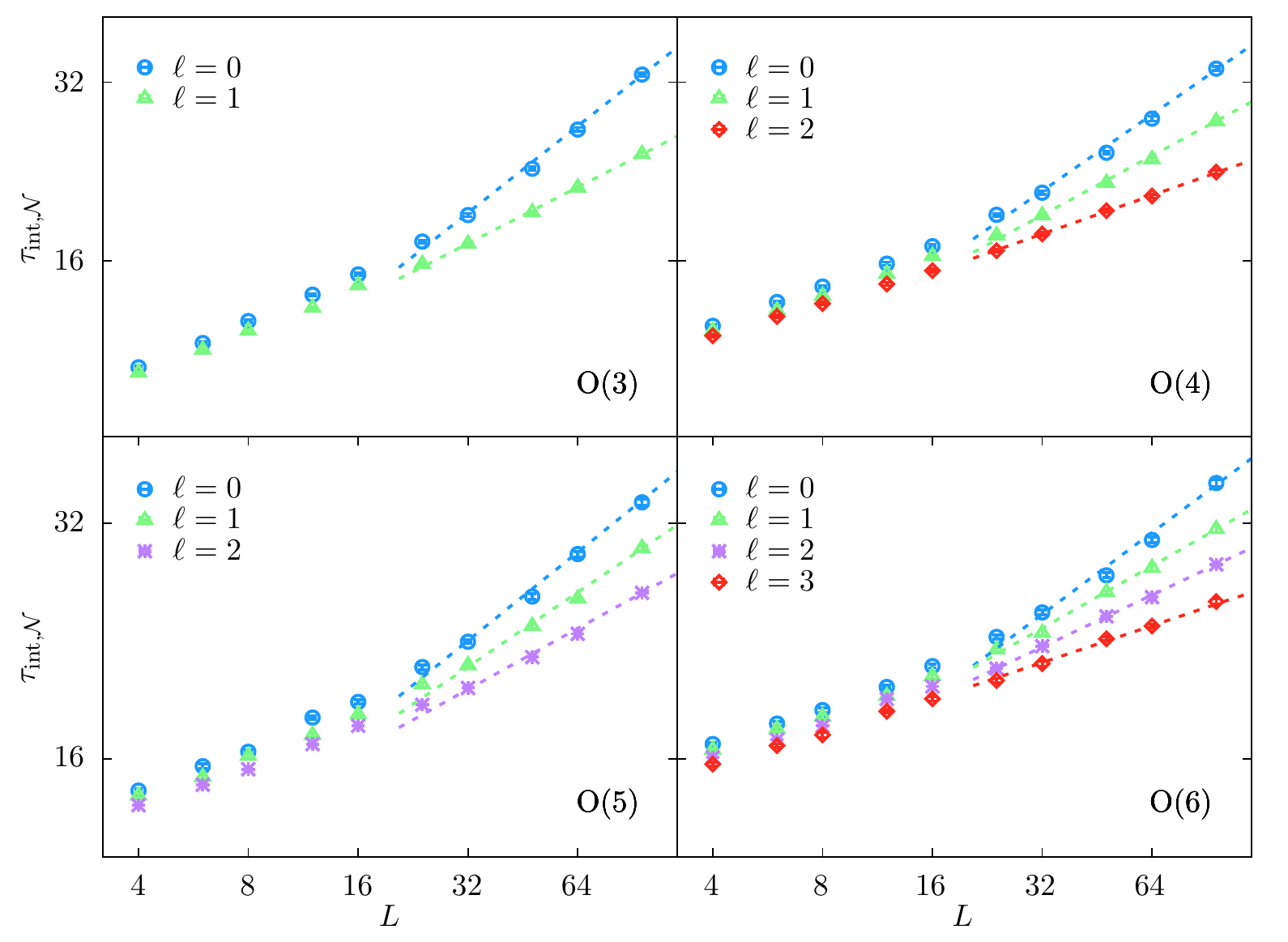}\\
	\caption
	{Integrated autocorrelation time $\tau_{\rm int,\mathcal{N}}$ at criticality for $N=3,4,5$ 
	and $6$ at different $\ell$ , in the unit of system sweep. 
	Points of different colors represent data for different $\ell$, and dashed lines here 
	have the same meaning with those in Fig.~\ref{fig:tauO2}.}
	\label{fig:tauO36}
\end{figure}

Utilizing the formula in Eq.~(\ref{eq:tau}) and the truncation strategy, we derive the integrated 
autocorrelation time $\tau_{\text{int},\mathcal{N}}$ at criticality. 
Based on the patterns observed in $\rho_{\mathcal{N}}(t)$, we select the windowing parameter $c$ 
to be $8$ for $N=2$, $\ell=1$, and $20$ for the remaining cases. 
Figure~\ref{fig:tauO2} presents the relationship between $\tau_{\text{int},\mathcal{N}}$ and 
system size for $N=2$, while Fig.~\ref{fig:tauO36} depicts the same for $N=3,4,5$ and $6$. 
Across all $N$ and $\ell$ values, $\tau_{\text{int},\mathcal{N}}$ appears to 
increase approximately as a power law with growing $L$, thereby indicating 
the manifestation of the critical slowing down phenomenon.

To quantify this, we fit the data for $\tau_{\text{int},\mathcal{N}}$ at larger $L$ values 
for all $N$ and $\ell$ using the following formula:
\begin{eqnarray} \label{eq:tau_int}
	\tau_{\text{int},\mathcal{N}}=AL^{z_{\text{int},\mathcal{N}}}.
\end{eqnarray}
The results of these fits are represented by the dashed lines in the corresponding figures, 
while the estimates for $z_{\text{int},\mathcal{N}}$ at different $N$ and $\ell$ are detailed in Tab.~\ref{tab:z}. 
The findings suggest that for each $N$, as $\ell$ increases, $z_{\text{int},\mathcal{N}}$ 
decreases monotonically, implying a higher efficiency of the worm algorithm at larger $\ell$. 
This observation can also be directly gleaned from the slopes of each line in Fig.~\ref{fig:tauO2} 
and \ref{fig:tauO36}.
Even for the worst case ($\ell=0$), the dynamic exponent $z$ is significantly smaller than 
that for the Metropolis algorithm, which has $z \approx 2.0$, meaning that the worm algorithms 
are much more efficient than the Metropolis method. 
In comparison with the Swendsen-Wang (SW) cluster method, which has $z =0.46(3)$ for 
the three-dimensional Ising model~\cite{Deng2007SW}, the worm algorithm for the worst case ($z \lesssim 0.5$)
seems to be comparable, and for the best case ($z \lesssim 0.3$) it seems to be somewhat 
more efficient, particularly for large $N$.
Taking into account the existence of `critical speeding-up' phemomenon 
for susceptibility-like quantities in worm-type simulations~\cite{Deng2007worm},
we conclude that the worm algorithm is at least as efficient as the SW cluster method. 

\section{Discussion}\label{sec:dis}

We present a successful reformulation of the O($N$) spin model into a series of graphical representations,
introduce corresponding worm algorithms to facilitate simulations and compare their dynamical critical behaviors.
It is observed that the algorithm corresponding to the representation with a larger $\ell$ demonstrates 
a smaller dynamical critical exponent, implying a superior level of efficiency. 
Additionally, we employ these graphical representations to derive corresponding tensor representations, 
which we believe will be useful for TNR or other numerical methods.

The availability of multiple graphical representations allows us to investigate the underlying mechanisms 
contributing to the superior efficiency of these algorithms. This exploration provides valuable insights 
into their effective operation.
To further enhance practical efficiency, our next step focuses on reducing the constant factor $A$ 
of the autocorrelation time. We propose achieving this improvement by incorporating irreversible 
Monte Carlo techniques, specifically the lifting 
technique~\cite{diaconis2000analysis,turitsyn2011irre,hu2017irre,elcci2018lifted}. 
Our preliminary test verifies the potential of this approach, motivating its adoption in our research.

In addition to providing a series of new graphical representations and corresponding algorithms, 
our work presents a general idea for reformulating systems describing vectors with multiple components 
into graphical representations. This involves breaking the vectors into the sum of analogs with 
fewer components, which are easier to represent graphically. The configurations can then be decomposed 
into a coupling of corresponding subgraphs, opening up prospects for future graphical representations 
of the CP($N-1$) spin model.

Moreover, this approach can be applied to overcome the complex action problem when solving the O($N$) 
spin model with a chemical potential $\mu$ coupled with conserved charges originating from 
the global O($N$) symmetry~\cite{bruckmann2015dual}. 
By binding the components related to the chemical potential as a copy of the XY spin vector 
during the decomposition of the total spin vector, the chemical potential term can be reduced 
to an extra weight factor that depends on the conserved charges in the graphical representation. 
In this new representation, the positive and negative currents for the chemical-potential-related XY 
spin vector are no longer conserved unless the extra integer variables defined on each site that 
stand for the conserved charges are taken into account. 
It is also noted that these graphical representations can provide 
a convenient and effective platform to study many physical problems 
like universal conductivity or halon physics~\cite{Chenkun2014,Chenkun2016,Chenkun2018}. 
Overall, our approach offers a versatile tool to reformulate complex systems in a more accessible 
and intuitive graphical form.

\begin{center}
{\bf ACKNOWLEDGMENTS}
\end{center}\par
We thank Nikolay Prokof'ev and Boris Svistunov for valuable discussions.
This work was supported by the National Natural Science Foundation of China (under Grant No.~12275263),
the Innovation Program for Quantum Science and Technology (under Grant No.~2021ZD0301900), 
and the National Key R\&D Program of China (under Grants No.~2018YFA0306501).\par

\appendix
\setcounter{equation}{0}
\renewcommand{\theequation}{\thesection.\arabic{equation}}

\section{Computation of Eq.~(\ref{eq:I})} \label{ap:measure}
According to the restriction $\sum_{\alpha}\mathcal{A}_{\alpha}^2+\sum_{\beta}\mathcal{B}_{\beta}^2=1$, 
if we rewrite $\mathcal{B}_{\beta}$ as $\mathcal{A}_{\ell+\beta}$ ($\beta=1,2,...,N-2\ell$), 
we can express $\mathcal{A}_{\alpha}$ in spherical coordinate system 
$(\theta_1,\theta_2,...,\theta_{N-\ell-1})$ as
\begin{equation}
\hspace{-10pt}\left\{
\begin{array}{ll}
\!\!\mathcal{A}_{1}          &=\!\cos\theta_1\\
\!\!\mathcal{A}_{2}          &=\!\sin\theta_1\cos\theta_2\\
\cdots&\\
\!\!\mathcal{A}_{N-\ell-1} \!\!\!\! &=\!\sin\theta_1\sin\theta_2\cdots\sin\theta_{N-\ell-2}\cos\theta_{N-\ell-1} \\
\!\!\mathcal{A}_{N-\ell}   &=\!\sin\theta_1\sin\theta_2\cdots\sin\theta_{N-\ell-2}\sin\theta_{N-\ell-1} ,
\end{array}\right.\!\!\!\!\!
\end{equation}
where $\theta_1,\theta_2,...,\theta_{N-\ell-1}\in[0,\pi/2]$.
It is standard that the surface element can be expressed as
\begin{align}
\prod_{i=1}^{N-\ell} \hspace{-3pt} d \mathcal{A}_{i}|_{\sum_{\alpha=1}^{N-\ell}\mathcal{A}_{\alpha}^2=1}\hspace{-3pt} =\hspace{-5pt}\prod_{i=1}^{N-\ell-2}\hspace{-5pt}\sin^{N-\ell-1-i}\theta_i \hspace{-5pt}\prod_{j=1}^{N-\ell-1}\hspace{-5pt} d \theta_j .
\end{align}
By using the identity that
\begin{equation}
\int_0^{\frac{\pi}{2}}\cos ^x \theta \sin ^y \theta d\theta=
\frac{\Gamma(\frac{x+1}{2})\Gamma(\frac{y+1}{2})}{2\Gamma(\frac{x+y+2}{2})} \; ,
\end{equation}
where  $\Gamma(\cdot)$ is the Gamma function, Eq.~(\ref{eq:I}) can be obtained directly.

\section{Acceptance Probabilities}\label{ap:acceptance_probabilities}
\noindent According to the Metropolis acceptance criterion, 
explicit expressions for $P_{\gamma\rightarrow \gamma'}^{\rm acc}(I\rightarrow I')
={\rm min}\{1,q^{(1)}\}$, $P_{\gamma,xy}^{\rm acc}(I\rightarrow I',sgn)=
{\rm min}\{1,q^{(2)}\}$ and $P_{\gamma,{\rm Is}}^{\rm acc}(I\rightarrow I')
={\rm min}\{1,q^{(3)}\}$ are as follows:\\ \\
(\uppercase\expandafter{\romannumeral1}) For $(\gamma,I)\rightarrow (\gamma',I')$:
%$P_{\gamma\rightarrow\gamma'}^{acc}$, $P_{\gamma,{xy}}^{acc}$ and $P_{\gamma,{\rm Is}}^{acc}$
\begin{equation}
	q^{(1)}=\frac{g_I+N}{g_I^{(\gamma)}+f(\gamma)}\frac{g_{I'}^{(\gamma')}+f(\gamma')}{g_{I'}+N}\; ,
\end{equation}
where
\begin{equation}
	f(\gamma)=\left\{\begin{array}{cc}
		2,& \text{ if $\gamma$ is an XY subgraph  }\\
		1,& \text{ if $\gamma$ is an Ising subgraph}
	\end{array}\right. \; ,
\end{equation}
\begin{equation}
	g_I^{(\gamma)}=\left\{\begin{array}{cc}
		k_I^{(\gamma)},& \text{ if $\gamma$ is an XY subgraph  }\\
		l_I^{(\gamma)},& \text{ if $\gamma$ is an Ising subgraph}
	\end{array}\right. \; ,
\end{equation}
and $g_I=\sum_{\alpha}k_I^{(\alpha)}+\sum_{\beta}l_I^{(\beta)}$. 
The expressions of $ k_I^{(\gamma)}$ and $l_I^{(\gamma)}$ are given by Eq.~(\ref{eq:kl}). 
\\\\
(\uppercase\expandafter{\romannumeral2}) When $\gamma$ is an XY subgraph: \\
(\romannumeral1) $m_{II'}^{sgn(\gamma)}\rightarrow m_{II'}^{sgn(\gamma)}+1$:
\begin{equation}
	q^{(2)}=\frac{K}{2(m_{II'}^{sgn(\gamma)}+1)}\frac{h_{I'}^{(\gamma)}+2}{h_{I'}+N}
\end{equation}
(\romannumeral2) $m_{II'}^{sgn(\gamma)}\rightarrow m_{II'}^{sgn(\gamma)}-1$:
\begin{equation}
	q^{(2)}=\frac{2m_{II'}^{sgn(\gamma)}}{K}\frac{h_{I}+N-2}{h_{I}^{(\gamma)}}.
\end{equation}\\
(\uppercase\expandafter{\romannumeral3}) When $\gamma$ is an Ising subgraph:\\
(\romannumeral1) $n_{II'}^{(\gamma)}\rightarrow n_{II'}^{(\gamma)}+1$:
\begin{equation}
	q^{(3)}=\frac{K}{n_{II'}^{(\gamma)}+1}\frac{h_{I'}^{(\gamma)}+1}{h_{I'}+N}
\end{equation}
(\romannumeral2) $n_{II'}^{(\gamma)}\rightarrow n_{II'}^{(\gamma)}-1$:
\begin{equation}
	q^{(3)}=\frac{n_{II'}^{(\gamma)}}{K}\frac{h_{I}+N-2}{h_{I}^{(\gamma)}-1}
\end{equation}
$h_i^{(\gamma)}$ and $h_i$ are defined as $h_i^{(\gamma)}=g_i^{(\gamma)}+\delta(i-I)+\delta(i-M)$ 
and $h_i=g_i+\delta(i-I)+\delta(i-M)$.

\vfill

\bibliographystyle{apsrev4-1}
\bibliography{reference}

\begin{thebibliography}{29}%
\makeatletter
\providecommand \@ifxundefined [1]{%
 \@ifx{#1\undefined}
}%
\providecommand \@ifnum [1]{%
 \ifnum #1\expandafter \@firstoftwo
 \else \expandafter \@secondoftwo
 \fi
}%
\providecommand \@ifx [1]{%
 \ifx #1\expandafter \@firstoftwo
 \else \expandafter \@secondoftwo
 \fi
}%
\providecommand \natexlab [1]{#1}%
\providecommand \enquote  [1]{``#1''}%
\providecommand \bibnamefont  [1]{#1}%
\providecommand \bibfnamefont [1]{#1}%
\providecommand \citenamefont [1]{#1}%
\providecommand \href@noop [0]{\@secondoftwo}%
\providecommand \href [0]{\begingroup \@sanitize@url \@href}%
\providecommand \@href[1]{\@@startlink{#1}\@@href}%
\providecommand \@@href[1]{\endgroup#1\@@endlink}%
\providecommand \@sanitize@url [0]{\catcode `\\12\catcode `\$12\catcode
  `\&12\catcode `\#12\catcode `\^12\catcode `\_12\catcode `\%12\relax}%
\providecommand \@@startlink[1]{}%
\providecommand \@@endlink[0]{}%
\providecommand \url  [0]{\begingroup\@sanitize@url \@url }%
\providecommand \@url [1]{\endgroup\@href {#1}{\urlprefix }}%
\providecommand \urlprefix  [0]{URL }%
\providecommand \Eprint [0]{\href }%
\providecommand \doibase [0]{http://dx.doi.org/}%
\providecommand \selectlanguage [0]{\@gobble}%
\providecommand \bibinfo  [0]{\@secondoftwo}%
\providecommand \bibfield  [0]{\@secondoftwo}%
\providecommand \translation [1]{[#1]}%
\providecommand \BibitemOpen [0]{}%
\providecommand \bibitemStop [0]{}%
\providecommand \bibitemNoStop [0]{.\EOS\space}%
\providecommand \EOS [0]{\spacefactor3000\relax}%
\providecommand \BibitemShut  [1]{\csname bibitem#1\endcsname}%
\let\auto@bib@innerbib\@empty
%</preamble>
\bibitem [{\citenamefont {Stanley}(1968)}]{stanley1968dependence}%
  \BibitemOpen
  \bibfield  {author} {\bibinfo {author} {\bibfnamefont {H.~E.}\ \bibnamefont
  {Stanley}},\ }\href@noop {} {\bibfield  {journal} {\bibinfo  {journal} {Phys.
  Rev. Lett.}\ }\textbf {\bibinfo {volume} {20}},\ \bibinfo {pages} {589}
  (\bibinfo {year} {1968})}\BibitemShut {NoStop}%
\bibitem [{\citenamefont {Binder}\ \emph {et~al.}(1986)\citenamefont {Binder},
  \citenamefont {Kalos}, \citenamefont {Hansen},\ and\ \citenamefont
  {Ceperley}}]{binder1986monte}%
  \BibitemOpen
  \bibfield  {author} {\bibinfo {author} {\bibfnamefont {K.}~\bibnamefont
  {Binder}}, \bibinfo {author} {\bibfnamefont {M.~H.}\ \bibnamefont {Kalos}},
  \bibinfo {author} {\bibfnamefont {J.}~\bibnamefont {Hansen}}, \ and\ \bibinfo
  {author} {\bibfnamefont {D.}~\bibnamefont {Ceperley}},\ }\href@noop {} {\emph
  {\bibinfo {title} {Monte Carlo methods in statistical physics}}}\ (\bibinfo
  {publisher} {Springer},\ \bibinfo {year} {1986})\BibitemShut {NoStop}%
\bibitem [{\citenamefont {Metropolis}\ \emph {et~al.}(1953)\citenamefont
  {Metropolis}, \citenamefont {Rosenbluth}, \citenamefont {Rosenbluth},
  \citenamefont {Teller},\ and\ \citenamefont
  {Teller}}]{metropolis1953equation}%
  \BibitemOpen
  \bibfield  {author} {\bibinfo {author} {\bibfnamefont {N.}~\bibnamefont
  {Metropolis}}, \bibinfo {author} {\bibfnamefont {A.~W.}\ \bibnamefont
  {Rosenbluth}}, \bibinfo {author} {\bibfnamefont {M.~N.}\ \bibnamefont
  {Rosenbluth}}, \bibinfo {author} {\bibfnamefont {A.~H.}\ \bibnamefont
  {Teller}}, \ and\ \bibinfo {author} {\bibfnamefont {E.}~\bibnamefont
  {Teller}},\ }\href@noop {} {\bibfield  {journal} {\bibinfo  {journal} {J.
  Chem. Phys.}\ }\textbf {\bibinfo {volume} {21}},\ \bibinfo {pages} {1087}
  (\bibinfo {year} {1953})}\BibitemShut {NoStop}%
\bibitem [{\citenamefont {Sokal}(1997)}]{sokal1997functional}%
  \BibitemOpen
  \bibfield  {author} {\bibinfo {author} {\bibfnamefont {A.}~\bibnamefont
  {Sokal}},\ }\href@noop {} {\enquote {\bibinfo {title} {Functional
  integration: Basics and applications},}\ } (\bibinfo {year}
  {1997})\BibitemShut {NoStop}%
\bibitem [{\citenamefont {Swendsen}\ and\ \citenamefont
  {Wang}(1987)}]{swendsen1987nonuniversal}%
  \BibitemOpen
  \bibfield  {author} {\bibinfo {author} {\bibfnamefont {R.~H.}\ \bibnamefont
  {Swendsen}}\ and\ \bibinfo {author} {\bibfnamefont {J.-S.}\ \bibnamefont
  {Wang}},\ }\href@noop {} {\bibfield  {journal} {\bibinfo  {journal} {Phys.
  Rev. Lett.}\ }\textbf {\bibinfo {volume} {58}},\ \bibinfo {pages} {86}
  (\bibinfo {year} {1987})}\BibitemShut {NoStop}%
\bibitem [{\citenamefont {Janke}(1996)}]{janke1996}%
  \BibitemOpen
  \bibfield  {author} {\bibinfo {author} {\bibfnamefont {W.}~\bibnamefont
  {Janke}},\ }\href {\doibase 10.1007/978-3-642-85238-1_3} {\emph {\bibinfo
  {title} {Computational Physics: Selected Methods Simple Exercises Serious
  Applications}}},\ edited by\ \bibinfo {editor} {\bibfnamefont {K.~H.}\
  \bibnamefont {Hoffmann}}\ and\ \bibinfo {editor} {\bibfnamefont
  {M.}~\bibnamefont {Schreiber}}\ (\bibinfo  {publisher} {Springer Berlin
  Heidelberg},\ \bibinfo {address} {Berlin, Heidelberg},\ \bibinfo {year}
  {1996})\ pp.\ \bibinfo {pages} {10--43}\BibitemShut {NoStop}%
\bibitem [{\citenamefont {Prokof'ev}\ and\ \citenamefont
  {Svistunov}(2001)}]{prokof2001worm}%
  \BibitemOpen
  \bibfield  {author} {\bibinfo {author} {\bibfnamefont {N.}~\bibnamefont
  {Prokof'ev}}\ and\ \bibinfo {author} {\bibfnamefont {B.}~\bibnamefont
  {Svistunov}},\ }\href@noop {} {\bibfield  {journal} {\bibinfo  {journal}
  {Phys. Rev. Lett.}\ }\textbf {\bibinfo {volume} {87}},\ \bibinfo {pages}
  {160601} (\bibinfo {year} {2001})}\BibitemShut {NoStop}%
\bibitem [{\citenamefont {Prokof'ev}\ \emph {et~al.}(1998)\citenamefont
  {Prokof'ev}, \citenamefont {Svistunov},\ and\ \citenamefont
  {Tupitsyn}}]{prokof1998worm}%
  \BibitemOpen
  \bibfield  {author} {\bibinfo {author} {\bibfnamefont {N.}~\bibnamefont
  {Prokof'ev}}, \bibinfo {author} {\bibfnamefont {B.}~\bibnamefont
  {Svistunov}}, \ and\ \bibinfo {author} {\bibfnamefont {I.}~\bibnamefont
  {Tupitsyn}},\ }\href@noop {} {\bibfield  {journal} {\bibinfo  {journal}
  {Phys. Lett. A}\ }\textbf {\bibinfo {volume} {238}},\ \bibinfo {pages} {253}
  (\bibinfo {year} {1998})}\BibitemShut {NoStop}%
\bibitem [{\citenamefont {Wolff}(atum)}]{wolff2010simulating}%
  \BibitemOpen
  \bibfield  {author} {\bibinfo {author} {\bibfnamefont {U.}~\bibnamefont
  {Wolff}},\ }\href@noop {} {\bibfield  {journal} {\bibinfo  {journal} {Nucl.
  Phys. B}\ }\textbf {\bibinfo {volume} {824}},\ \bibinfo {pages} {254}
  (\bibinfo {year} {2010); U. Wolff, Nucl. Phys. B \textbf{834}, 395
  (2010)(Erratum})}\BibitemShut {NoStop}%
\bibitem [{\citenamefont {Bruckmann}\ \emph {et~al.}(2015)\citenamefont
  {Bruckmann}, \citenamefont {Gattringer}, \citenamefont {Kloiber},\ and\
  \citenamefont {Sulejmanpasic}}]{bruckmann2015dual}%
  \BibitemOpen
  \bibfield  {author} {\bibinfo {author} {\bibfnamefont {F.}~\bibnamefont
  {Bruckmann}}, \bibinfo {author} {\bibfnamefont {C.}~\bibnamefont
  {Gattringer}}, \bibinfo {author} {\bibfnamefont {T.}~\bibnamefont {Kloiber}},
  \ and\ \bibinfo {author} {\bibfnamefont {T.}~\bibnamefont {Sulejmanpasic}},\
  }\href@noop {} {\bibfield  {journal} {\bibinfo  {journal} {Phys. Lett. B}\
  }\textbf {\bibinfo {volume} {749}},\ \bibinfo {pages} {495} (\bibinfo {year}
  {2015})}\BibitemShut {NoStop}%
\bibitem [{\citenamefont {Katz}\ \emph {et~al.}(2017)\citenamefont {Katz},
  \citenamefont {Niedermayer}, \citenamefont {N{\'o}gr{\'a}di},\ and\
  \citenamefont {T{\"o}r{\"o}k}}]{katz2017comparison}%
  \BibitemOpen
  \bibfield  {author} {\bibinfo {author} {\bibfnamefont {S.~D.}\ \bibnamefont
  {Katz}}, \bibinfo {author} {\bibfnamefont {F.}~\bibnamefont {Niedermayer}},
  \bibinfo {author} {\bibfnamefont {D.}~\bibnamefont {N{\'o}gr{\'a}di}}, \ and\
  \bibinfo {author} {\bibfnamefont {C.}~\bibnamefont {T{\"o}r{\"o}k}},\
  }\href@noop {} {\bibfield  {journal} {\bibinfo  {journal} {Phys. Rev. D}\
  }\textbf {\bibinfo {volume} {95}},\ \bibinfo {pages} {054506} (\bibinfo
  {year} {2017})}\BibitemShut {NoStop}%
\bibitem [{\citenamefont {Deng}\ \emph
  {et~al.}(2007{\natexlab{a}})\citenamefont {Deng}, \citenamefont {Garoni},
  \citenamefont {Machta}, \citenamefont {Ossola}, \citenamefont {Polin},\ and\
  \citenamefont {Sokal}}]{Deng2007SW}%
  \BibitemOpen
  \bibfield  {author} {\bibinfo {author} {\bibfnamefont {Y.}~\bibnamefont
  {Deng}}, \bibinfo {author} {\bibfnamefont {T.~M.}\ \bibnamefont {Garoni}},
  \bibinfo {author} {\bibfnamefont {J.}~\bibnamefont {Machta}}, \bibinfo
  {author} {\bibfnamefont {G.}~\bibnamefont {Ossola}}, \bibinfo {author}
  {\bibfnamefont {M.}~\bibnamefont {Polin}}, \ and\ \bibinfo {author}
  {\bibfnamefont {A.~D.}\ \bibnamefont {Sokal}},\ }\href {\doibase
  10.1103/PhysRevLett.99.055701} {\bibfield  {journal} {\bibinfo  {journal}
  {Phys. Rev. Lett.}\ }\textbf {\bibinfo {volume} {99}},\ \bibinfo {pages}
  {055701} (\bibinfo {year} {2007}{\natexlab{a}})}\BibitemShut {NoStop}%
\bibitem [{\citenamefont {Evenbly}\ and\ \citenamefont
  {Vidal}(2015)}]{evenbly2015}%
  \BibitemOpen
  \bibfield  {author} {\bibinfo {author} {\bibfnamefont {G.}~\bibnamefont
  {Evenbly}}\ and\ \bibinfo {author} {\bibfnamefont {G.}~\bibnamefont
  {Vidal}},\ }\href {\doibase 10.1103/PhysRevLett.115.180405} {\bibfield
  {journal} {\bibinfo  {journal} {Phys. Rev. Lett.}\ }\textbf {\bibinfo
  {volume} {115}},\ \bibinfo {pages} {180405} (\bibinfo {year}
  {2015})}\BibitemShut {NoStop}%
\bibitem [{\citenamefont {Liu}\ \emph {et~al.}(2013)\citenamefont {Liu},
  \citenamefont {Meurice}, \citenamefont {Qin}, \citenamefont {Unmuth-Yockey},
  \citenamefont {Xiang}, \citenamefont {Xie}, \citenamefont {Yu},\ and\
  \citenamefont {Zou}}]{liuyuzhi2013}%
  \BibitemOpen
  \bibfield  {author} {\bibinfo {author} {\bibfnamefont {Y.}~\bibnamefont
  {Liu}}, \bibinfo {author} {\bibfnamefont {Y.}~\bibnamefont {Meurice}},
  \bibinfo {author} {\bibfnamefont {M.~P.}\ \bibnamefont {Qin}}, \bibinfo
  {author} {\bibfnamefont {J.}~\bibnamefont {Unmuth-Yockey}}, \bibinfo {author}
  {\bibfnamefont {T.}~\bibnamefont {Xiang}}, \bibinfo {author} {\bibfnamefont
  {Z.~Y.}\ \bibnamefont {Xie}}, \bibinfo {author} {\bibfnamefont {J.~F.}\
  \bibnamefont {Yu}}, \ and\ \bibinfo {author} {\bibfnamefont {H.}~\bibnamefont
  {Zou}},\ }\href {\doibase 10.1103/PhysRevD.88.056005} {\bibfield  {journal}
  {\bibinfo  {journal} {Phys. Rev. D}\ }\textbf {\bibinfo {volume} {88}},\
  \bibinfo {pages} {056005} (\bibinfo {year} {2013})}\BibitemShut {NoStop}%
\bibitem [{\citenamefont {Madras}\ and\ \citenamefont
  {Sokal}(1988)}]{madras1988pivot}%
  \BibitemOpen
  \bibfield  {author} {\bibinfo {author} {\bibfnamefont {N.}~\bibnamefont
  {Madras}}\ and\ \bibinfo {author} {\bibfnamefont {A.~D.}\ \bibnamefont
  {Sokal}},\ }\href@noop {} {\bibfield  {journal} {\bibinfo  {journal} {J.
  Stat. Phys.}\ }\textbf {\bibinfo {volume} {50}},\ \bibinfo {pages} {109}
  (\bibinfo {year} {1988})}\BibitemShut {NoStop}%
\bibitem [{\citenamefont {Xu}\ \emph {et~al.}(2019)\citenamefont {Xu},
  \citenamefont {Sun}, \citenamefont {Jian-Ping},\ and\ \citenamefont
  {Deng}}]{Xu2019}%
  \BibitemOpen
  \bibfield  {author} {\bibinfo {author} {\bibfnamefont {W.}~\bibnamefont
  {Xu}}, \bibinfo {author} {\bibfnamefont {Y.}~\bibnamefont {Sun}}, \bibinfo
  {author} {\bibfnamefont {L.}~\bibnamefont {Jian-Ping}}, \ and\ \bibinfo
  {author} {\bibfnamefont {Y.}~\bibnamefont {Deng}},\ }\href {\doibase
  10.1103/PhysRevB.100.064525} {\bibfield  {journal} {\bibinfo  {journal}
  {Phys. Rev. B}\ }\textbf {\bibinfo {volume} {100}},\ \bibinfo {pages}
  {064525} (\bibinfo {year} {2019})}\BibitemShut {NoStop}%
\bibitem [{\citenamefont {Deng}\ \emph {et~al.}(2005)\citenamefont {Deng},
  \citenamefont {Bl{\"o}te},\ and\ \citenamefont
  {Nightingale}}]{deng2005surface}%
  \BibitemOpen
  \bibfield  {author} {\bibinfo {author} {\bibfnamefont {Y.}~\bibnamefont
  {Deng}}, \bibinfo {author} {\bibfnamefont {H.~W.}\ \bibnamefont {Bl{\"o}te}},
  \ and\ \bibinfo {author} {\bibfnamefont {M.}~\bibnamefont {Nightingale}},\
  }\href@noop {} {\bibfield  {journal} {\bibinfo  {journal} {Phys. Rev. E}\
  }\textbf {\bibinfo {volume} {72}},\ \bibinfo {pages} {016128} (\bibinfo
  {year} {2005})}\BibitemShut {NoStop}%
\bibitem [{\citenamefont {Deng}(2006)}]{deng2006bulk}%
  \BibitemOpen
  \bibfield  {author} {\bibinfo {author} {\bibfnamefont {Y.}~\bibnamefont
  {Deng}},\ }\href@noop {} {\bibfield  {journal} {\bibinfo  {journal} {Phys.
  Rev. E}\ }\textbf {\bibinfo {volume} {73}},\ \bibinfo {pages} {056116}
  (\bibinfo {year} {2006})}\BibitemShut {NoStop}%
\bibitem [{\citenamefont {Fernandez}\ \emph {et~al.}(2005)\citenamefont
  {Fernandez}, \citenamefont {Mart{\'\i}n-Mayor}, \citenamefont {Sciretti},
  \citenamefont {Taranc{\'o}n},\ and\ \citenamefont
  {Velasco}}]{fernandez2005O5}%
  \BibitemOpen
  \bibfield  {author} {\bibinfo {author} {\bibfnamefont {L.}~\bibnamefont
  {Fernandez}}, \bibinfo {author} {\bibfnamefont {V.}~\bibnamefont
  {Mart{\'\i}n-Mayor}}, \bibinfo {author} {\bibfnamefont {D.}~\bibnamefont
  {Sciretti}}, \bibinfo {author} {\bibfnamefont {A.}~\bibnamefont
  {Taranc{\'o}n}}, \ and\ \bibinfo {author} {\bibfnamefont {J.}~\bibnamefont
  {Velasco}},\ }\href@noop {} {\bibfield  {journal} {\bibinfo  {journal} {Phys.
  Lett. B}\ }\textbf {\bibinfo {volume} {628}},\ \bibinfo {pages} {281}
  (\bibinfo {year} {2005})}\BibitemShut {NoStop}%
\bibitem [{jia()}]{jianpinglv_private_communication}%
  \BibitemOpen
  \href@noop {} {}\bibinfo {howpublished} {Private communication with Jian-Ping
  Lv at Anhui Normal University, Wuhu, Anhui, China. This estimate aligns with
  that of Ref.~\cite{loison199906}, but offers a higher degree of
  accuracy.}\BibitemShut {Stop}%
\bibitem [{\citenamefont {Deng}\ \emph
  {et~al.}(2007{\natexlab{b}})\citenamefont {Deng}, \citenamefont {Garoni},\
  and\ \citenamefont {Sokal}}]{Deng2007worm}%
  \BibitemOpen
  \bibfield  {author} {\bibinfo {author} {\bibfnamefont {Y.}~\bibnamefont
  {Deng}}, \bibinfo {author} {\bibfnamefont {T.~M.}\ \bibnamefont {Garoni}}, \
  and\ \bibinfo {author} {\bibfnamefont {A.~D.}\ \bibnamefont {Sokal}},\ }\href
  {\doibase 10.1103/PhysRevLett.99.110601} {\bibfield  {journal} {\bibinfo
  {journal} {Phys. Rev. Lett.}\ }\textbf {\bibinfo {volume} {99}},\ \bibinfo
  {pages} {110601} (\bibinfo {year} {2007}{\natexlab{b}})}\BibitemShut
  {NoStop}%
\bibitem [{\citenamefont {Diaconis}\ \emph {et~al.}(2000)\citenamefont
  {Diaconis}, \citenamefont {Holmes},\ and\ \citenamefont
  {Neal}}]{diaconis2000analysis}%
  \BibitemOpen
  \bibfield  {author} {\bibinfo {author} {\bibfnamefont {P.}~\bibnamefont
  {Diaconis}}, \bibinfo {author} {\bibfnamefont {S.}~\bibnamefont {Holmes}}, \
  and\ \bibinfo {author} {\bibfnamefont {R.~M.}\ \bibnamefont {Neal}},\
  }\href@noop {} {\bibfield  {journal} {\bibinfo  {journal} {Ann. Appl.
  Probab}\ ,\ \bibinfo {pages} {726}} (\bibinfo {year} {2000})}\BibitemShut
  {NoStop}%
\bibitem [{\citenamefont {Turitsyn}\ \emph {et~al.}(2011)\citenamefont
  {Turitsyn}, \citenamefont {Chertkov},\ and\ \citenamefont
  {Vucelja}}]{turitsyn2011irre}%
  \BibitemOpen
  \bibfield  {author} {\bibinfo {author} {\bibfnamefont {K.~S.}\ \bibnamefont
  {Turitsyn}}, \bibinfo {author} {\bibfnamefont {M.}~\bibnamefont {Chertkov}},
  \ and\ \bibinfo {author} {\bibfnamefont {M.}~\bibnamefont {Vucelja}},\
  }\href@noop {} {\bibfield  {journal} {\bibinfo  {journal} {Physica D}\
  }\textbf {\bibinfo {volume} {240}},\ \bibinfo {pages} {410} (\bibinfo {year}
  {2011})}\BibitemShut {NoStop}%
\bibitem [{\citenamefont {Hu}\ \emph {et~al.}(2017)\citenamefont {Hu},
  \citenamefont {Chen},\ and\ \citenamefont {Deng}}]{hu2017irre}%
  \BibitemOpen
  \bibfield  {author} {\bibinfo {author} {\bibfnamefont {H.}~\bibnamefont
  {Hu}}, \bibinfo {author} {\bibfnamefont {X.}~\bibnamefont {Chen}}, \ and\
  \bibinfo {author} {\bibfnamefont {Y.}~\bibnamefont {Deng}},\ }\href@noop {}
  {\bibfield  {journal} {\bibinfo  {journal} {Front. Phys.}\ }\textbf {\bibinfo
  {volume} {12}},\ \bibinfo {pages} {120503} (\bibinfo {year}
  {2017})}\BibitemShut {NoStop}%
\bibitem [{\citenamefont {El{\c{c}}i}\ \emph {et~al.}(2018)\citenamefont
  {El{\c{c}}i}, \citenamefont {Grimm}, \citenamefont {Ding}, \citenamefont
  {Nasrawi}, \citenamefont {Garoni},\ and\ \citenamefont
  {Deng}}]{elcci2018lifted}%
  \BibitemOpen
  \bibfield  {author} {\bibinfo {author} {\bibfnamefont {E.~M.}\ \bibnamefont
  {El{\c{c}}i}}, \bibinfo {author} {\bibfnamefont {J.}~\bibnamefont {Grimm}},
  \bibinfo {author} {\bibfnamefont {L.}~\bibnamefont {Ding}}, \bibinfo {author}
  {\bibfnamefont {A.}~\bibnamefont {Nasrawi}}, \bibinfo {author} {\bibfnamefont
  {T.~M.}\ \bibnamefont {Garoni}}, \ and\ \bibinfo {author} {\bibfnamefont
  {Y.}~\bibnamefont {Deng}},\ }\href@noop {} {\bibfield  {journal} {\bibinfo
  {journal} {Phys. Rev. E}\ }\textbf {\bibinfo {volume} {97}},\ \bibinfo
  {pages} {042126} (\bibinfo {year} {2018})}\BibitemShut {NoStop}%
\bibitem [{\citenamefont {Chen}\ \emph {et~al.}(2014)\citenamefont {Chen},
  \citenamefont {Liu}, \citenamefont {Deng}, \citenamefont {Pollet},\ and\
  \citenamefont {Nikolay}}]{Chenkun2014}%
  \BibitemOpen
  \bibfield  {author} {\bibinfo {author} {\bibfnamefont {K.}~\bibnamefont
  {Chen}}, \bibinfo {author} {\bibfnamefont {L.}~\bibnamefont {Liu}}, \bibinfo
  {author} {\bibfnamefont {Y.}~\bibnamefont {Deng}}, \bibinfo {author}
  {\bibfnamefont {L.}~\bibnamefont {Pollet}}, \ and\ \bibinfo {author}
  {\bibfnamefont {P.}~\bibnamefont {Nikolay}},\ }\href {\doibase
  10.1103/PhysRevLett.112.030402} {\bibfield  {journal} {\bibinfo  {journal}
  {Phys. Rev. Lett.}\ }\textbf {\bibinfo {volume} {112}},\ \bibinfo {pages}
  {030402} (\bibinfo {year} {2014})}\BibitemShut {NoStop}%
\bibitem [{\citenamefont {Huang}\ \emph {et~al.}(2016)\citenamefont {Huang},
  \citenamefont {Chen}, \citenamefont {Deng},\ and\ \citenamefont
  {Svistunov}}]{Chenkun2016}%
  \BibitemOpen
  \bibfield  {author} {\bibinfo {author} {\bibfnamefont {Y.}~\bibnamefont
  {Huang}}, \bibinfo {author} {\bibfnamefont {K.}~\bibnamefont {Chen}},
  \bibinfo {author} {\bibfnamefont {Y.}~\bibnamefont {Deng}}, \ and\ \bibinfo
  {author} {\bibfnamefont {B.}~\bibnamefont {Svistunov}},\ }\href {\doibase
  10.1103/PhysRevB.94.220502} {\bibfield  {journal} {\bibinfo  {journal} {Phys.
  Rev. B}\ }\textbf {\bibinfo {volume} {94}},\ \bibinfo {pages} {220502}
  (\bibinfo {year} {2016})}\BibitemShut {NoStop}%
\bibitem [{\citenamefont {Chen}\ \emph {et~al.}(2018)\citenamefont {Chen},
  \citenamefont {Huang}, \citenamefont {Deng},\ and\ \citenamefont
  {Svistunov}}]{Chenkun2018}%
  \BibitemOpen
  \bibfield  {author} {\bibinfo {author} {\bibfnamefont {K.}~\bibnamefont
  {Chen}}, \bibinfo {author} {\bibfnamefont {Y.}~\bibnamefont {Huang}},
  \bibinfo {author} {\bibfnamefont {Y.}~\bibnamefont {Deng}}, \ and\ \bibinfo
  {author} {\bibfnamefont {B.}~\bibnamefont {Svistunov}},\ }\href {\doibase
  10.1103/PhysRevB.98.214516} {\bibfield  {journal} {\bibinfo  {journal} {Phys.
  Rev. B}\ }\textbf {\bibinfo {volume} {98}},\ \bibinfo {pages} {214516}
  (\bibinfo {year} {2018})}\BibitemShut {NoStop}%
\bibitem [{\citenamefont {Loison}(1999)}]{loison199906}%
  \BibitemOpen
  \bibfield  {author} {\bibinfo {author} {\bibfnamefont {D.}~\bibnamefont
  {Loison}},\ }\href@noop {} {\bibfield  {journal} {\bibinfo  {journal}
  {Physica A}\ }\textbf {\bibinfo {volume} {271}},\ \bibinfo {pages} {157}
  (\bibinfo {year} {1999})}\BibitemShut {NoStop}%
\end{thebibliography}%

\end{document}